\documentclass[12pt]{iopart}
\usepackage{slashed}
\usepackage{iopams}

\begin{document}

\newcommand\p{\partial}
\newcommand\Psibar{\overline{\Psi}}
\newcommand\psibar{\overline{\psi}}
\newcommand\rmc{\mathrm{c}}
\newcommand\rmic{\mathrm{ic}}
\newcommand\rmT{\mathrm{T}}
\newcommand\bW{\boldsymbol{W}}
\newcommand\Sdual{{}^{*}\!S}
\newcommand\sdual{{}^{*}\!s}
\newcommand\spart{\slashed{\partial}}
\newcommand\lspart{\overleftarrow{\slashed{\partial}}}

\title{Algebraic inversion of the Dirac equation for the vector potential in the non-abelian case}

\author{S M Inglis and P D Jarvis}

\address{School of Mathematics and Physics, University of Tasmania, Sandy Bay Campus, Private Bag 37, Hobart, Tasmania, Australia 7001}

\eads{\mailto{sminglis@utas.edu.au}, \mailto{Peter.Jarvis@utas.edu.au}}

\begin{abstract}
We study the Dirac equation for spinor wavefunctions minimally coupled to an external field, from the perspective of an algebraic system of linear equations for the vector potential. By analogy with the method in electromagnetism, which has been well-studied, and leads to classical solutions of the Maxwell-Dirac equations, we set up the formalism for non-abelian gauge symmetry, with the $SU(2)$ group and the case of four-spinor doublets. An extended isospin-charge conjugation operator is defined, enabling the hermiticity constraint on the gauge potential to be imposed in a covariant fashion, and rendering the algebraic system tractable. The outcome is an invertible linear equation for the non-abelian vector potential in terms of bispinor current densities. We show that, via application of suitable extended Fierz identities, the solution of this system for the non-abelian vector potential is a rational expression involving only Pauli scalar and Pauli triplet, Lorentz scalar, vector and axial vector current densities, albeit in the non-closed form of a Neumann series.
\end{abstract}


\section{Introduction}
One of the central equations of electrodynamics is the Dirac equation, the spinor solutions of which describe the states of a free relativistic spin-$\frac{1}{2}$ particle, such as an electron. The electromagnetic vector potential $A_{\mu}$ can be introduced by imposing local $U(1)$ gauge covariance, and since elements of this group commute, the electromagnetic interaction is an \textit{abelian} gauge field. In order to describe the way in which an electron interacts with its own self-consistent electromagnetic field, one approach to solutions of the coupled Maxwell-Dirac equations is to ``invert'' the $U(1)$ gauge covariant Dirac equation to solve for $A_{\mu}$ in terms of $\psi$, then to substitute this into the Maxwell equations, which have the Dirac current as the source term \cite{Booth-2000}. The full Maxwell-Dirac system is very complicated, and no exact, closed form solutions have yet been discovered, although global solutions have been shown to exist \cite{Flato-Simon-Taflin-1997}. Solutions for certain simplified cases using the algebraic inversion technique have been derived by Radford and Booth, with restrictions such as a static Dirac field with further assumptions of spherical (Coulomb) \cite{Radford-1996} and cylindrical (charged wire) symmetry \cite{Booth-Radford-1997}. It is interesting to note further that a spherically symmetric external Coulomb field requires the existence of a magnetic monopole term in the vector potential \cite{Radford-1996}, \cite{Radford-Booth-1999}. Another approach for finding solutions to the Maxwell-Dirac system has been pursued by Legg \cite{Legg-2007}, where the algebraic inversion was handled without fixing the gauge. The actions of various simply transitive subgroups of the Poincar\'{e} group were then applied in order to reduce the system of equations to a set of algebraic relations, involving the group invariant components of a general gauge invariant vector field. A global, closed-form solution was discovered for one of these subgroups, which exhibited unbounded laminar stream flow of the relativistic electron four-current along the z-axis, with mutually perpendicular electric and magnetic fields, \cite{Legg-2007}.

With regards to the inversion of the Dirac equation itself, in an early study \cite{Eliezer-1958} Eliezer pointed out that, when writing the Dirac equation in the form
\begin{equation}
MA_{\mu}=R_{\mu}
\end{equation}
where $M$ contains only Dirac spinor components, because $\det M=0$, the four equations are not linearly independent, and so $M$ is not invertible. However, a more recent study by Booth, Legg and Jarvis \cite{Booth-Legg-Jarvis-2001} demonstrated that the system of linear equations is indeed invertible if strictly real solutions of the vector potential are required. In the current study, we have assumed this condition holds for the $SU(2)$ non-abelian case.

A major publication by Takabayasi \cite{Takabayasi-1957} explains that by writing the Dirac equation in terms of Lorentz scalars, pseudoscalars, vectors and axial vectors, the model becomes analogous to a relativistic hydrodynamical one, and the current solutions describe fields of relativistic charged flow. Non-abelian fluid flow was investigated in \cite{Bistrovic-2003}, where an application to a quark-gluon plasma, involving the $SU(3)$ gauge group, was discussed.

In the current study, we are attempting to derive an inverted form of the Dirac equation, for interactions involving the $SU(2)$ gauge group. Section 2 contains a brief summary of the inversion for the abelian case, outlining the method used in \cite{Booth-Legg-Jarvis-2001}, where the regular and charge conjugate versions of the Dirac equations are multiplied by $\psibar\gamma_{\mu}$ and $\psibar{}^{\rmc}\gamma_{\mu}$ respectively, allowing one to work in terms of bispinor current densities, such as $\psibar\gamma_{\mu}\psi$. The rank-2 skew tensors are then eliminated via subtraction and the utilization of charge conjugate bispinor identities for the semi-classical case of commuting wavefunctions. The well-known inverted form is then easily achieved by inverting the remaining Lorentz scalar. An alternative inversion, obtained as a result of multiplication by $\psibar\gamma_{5}\gamma_{\mu}$ is also given, as well as several consistency conditions.

Section 3 sets up the inversion of the non-abelian Dirac equation for the $SU(2)$ gauge field $W_{a\mu}$, and follows the same path of logic as the abelian inversion, where we primarily focus on forming bispinors with an extra doublet degree of freedom, such as $\Psibar\tau_{a}\gamma_{\mu}\Psi$, by multiplying by $\Psibar\tau_{a}\gamma_{\mu}$. An extended form of the charge conjugate equation is required in order make the form of the gauge potential covariant, dubbed the ``isospin-charge conjugation''. This method turns out to yield a much more difficult inversion problem compared with the abelian case, involving the expression of an inverse matrix of form $(I-N)^{-1}$ by a Neumann series. Conditions of convergence are stated, but not verified for this case, and a closed form of the inverse matrix has not been obtained as of yet. We end the section with a brief comment on alternative bispinor-forming multiplication options, which yield couplings of various non-abelian current densities to $W_{a\mu}$. A full list of coupling equations is deffered to appendix B.

In section 4 the Fierz expansion of the $\Psi\Psibar$ term is given with the motivation of developing Fierz identities, describing relationships between various non-abelian current densities. In particular, general Fierz identities for non-abelian Lorentz vector current and axial current density products, $J_{i\mu}K_{j\nu}$, are derived in appendix C using this expansion. Using the abelian case \cite{Crawford-1985} as a guide, various antisymmetric combinations of these identities are then constructed to produce expressions for the rank-2 skew tensor current density $S_{i\mu\nu}$ and its dual $\Sdual_{j\mu\nu}$, solely in terms of Lorentz scalars and vector combinations. Pauli singlet $(i=0)$ and vector triplet $(i=1,2,3)$ are treated case-wise, due to their different algebraic behaviours. The rank-2 skew tensor current densities may then be eliminated from the expression for the inverse matrix $(I-N)^{-1}$ entirely.

Conclusions and prospects for future study are given in section 5, followed by a robust appendix. The first two appendix sections provide lists of commonly used Pauli and Dirac matrix identities, as well as a list of field-current density coupling equations, such as those discussed in section 2, but for a greater variety of multiplication options and algebraic combinations. The last two sections of the appendix include more detailed derivations of the Fierz identities discussed in section 4.

\section{The abelian inversion case}
The Dirac equation for a spin-$\frac{1}{2}$ fermion of charge $q$ moving under the influence of an electromagnetic field is
\begin{equation}
(\rmi\slashed{\partial}-q\slashed{A}-m)\psi=0
\end{equation}
which is covariant under a $U(1)$ (abelian) gauge transformation. Here we invoke the Feynman slash notation $\slashed{a}\equiv\gamma^{\nu}a_{\nu}$. Rearranging into a more convenient form
\begin{equation}\label{Emag Dirac eqn compressed}
\gamma^{\nu}\psi A_{\nu}=\phi
\end{equation}
where $\phi\equiv q^{-1}(\rmi\slashed{\partial}-m)\psi$, we can obtain an expression in terms of Dirac bispinors by (\ref{Emag Dirac eqn compressed}) by $\psibar\gamma_{\mu}$ to form bispinors and applying the identity $\gamma_{\mu}\gamma^{\nu}=\delta_{\mu}{}^{\nu}-\rmi\sigma_{\mu}{}^{\nu}$, such that
\begin{equation}\label{U(1) Dirac eqn after pre-mult}
\psibar\psi A_{\mu}-\rmi\psibar\sigma_{\mu}{}^{\nu}\psi A_{\nu}=\psibar\gamma_{\mu}\phi.
\end{equation}
The form of the charge conjugate Dirac equation is the same, but with the sign of the charge in $\phi^{\rmc}$ reversed. Note that we define the charge conjugate spinor as usual as \cite{Itzykson-Zuber-1980}
\begin{equation}
\psi^{\rmc}=C\psibar{}^{\rmT}=\rmi\gamma^{2}\gamma^{0}\psibar{}^{\rmT}.
\end{equation}
With this in mind, we can define a relationship between bispinors involving charge conjugate spinors and regular bispinors
\begin{equation}\label{U(1) bispinor relation}
\psibar{}^{\rmc}\Gamma\psi^{\rmc}=-\psibar C^{-1}\Gamma^{\rmT}C\psi
\end{equation}
for any $\Gamma$ in the Dirac algebra. In order to perform the inversion, we apply the charge conjugation identities
\numparts
\begin{eqnarray}
C^{-1}\gamma_{\mu}^{\rmT}C=-\gamma_{\mu} \\
C^{-1}\sigma_{\mu\nu}^{\rmT}C=-\sigma_{\mu\nu} \\
C^{-1}(\gamma_{\mu}\gamma_{\nu})^{\rmT}C=\gamma_{\nu}\gamma_{\mu}
\end{eqnarray}
\endnumparts
to the appropriate bispinor current densities $\psibar\Gamma\psi$, to obtain the relationships between the bispinor densities appearing in (\ref{U(1) Dirac eqn after pre-mult}) and their charge conjugates, which written explicitly are
\numparts
\begin{eqnarray}
\psibar{}^{\rmc}\psi^{\rmc}=-\psibar\psi \label{U(1) bispinor relation Gamma=I} \\
\psibar{}^{\rmc}\gamma_{\mu}\psi^{\rmc}=\psibar\gamma_{\mu}\psi \\
\psibar{}^{\rmc}\sigma_{\mu\nu}\psi^{\rmc}=\psibar\sigma_{\mu\nu}\psi \\
\psibar{}^{\rmc}\gamma_{\mu}\slashed{\partial}\psi^{\rmc}=-\psibar\overleftarrow{\slashed{\partial}}\gamma_{\mu}\psi. \label{U(1) bispinor relation Gamma=gamma-mu partial}
\end{eqnarray}
\endnumparts
Note that the allocation of signs would be different if we were to treat the spinors $\psi$ as field operators $\widehat{\psi}$ instead of semi-classical wavefunctions, however the quantum field aspects of the Dirac equation are beyond the scope of this paper. Now, subtracting the charge conjugate of (\ref{U(1) Dirac eqn after pre-mult}) from (\ref{U(1) Dirac eqn after pre-mult}) itself, applying the identities (\ref{U(1) bispinor relation Gamma=I})-(\ref{U(1) bispinor relation Gamma=gamma-mu partial}) and rearranging, we obtain the well-known inverted form of the $U(1)$ covariant Dirac equation
\begin{equation}
A_{\mu}=\frac{1}{2q}\frac{\rmi(\psibar\gamma_{\mu}\slashed{\partial}\psi-\psibar\overleftarrow{\slashed{\partial}}\gamma_{\mu}\psi)-2mj_{\mu}}{\psibar\psi}
\end{equation}
where $j_{\mu}\equiv\psibar\gamma_{\mu}\psi$, the probability current density. There is an alternative inverted form, which involves multiplying (\ref{Emag Dirac eqn compressed}) by $\psibar\gamma_{5}\gamma_{\mu}$, in which case the mass term vanishes
\begin{equation}\label{Abelian alternate inverted form}
A_{\mu}=\frac{\rmi}{2q}\frac{\psibar\gamma_{5}\gamma_{\mu}\slashed{\partial}\psi+\psibar\overleftarrow{\slashed{\partial}}\gamma_{5}\gamma_{\mu}\psi}{\psibar\gamma_{5}\psi}
\end{equation}
where $\gamma_{5}\equiv\rmi\gamma_{0}\gamma_{1}\gamma_{2}\gamma_{3}$. By considering spinors that have zero real inner product with $\gamma_{\mu}\psi$, we can obtain the following consistency conditions \cite{Booth-Legg-Jarvis-2001}
\begin{eqnarray}
\partial_{\mu}j^{\mu}=0 \\
\partial_{\mu}k^{\mu}=-2\rmi m\psibar\gamma_{5}\psi \\
\psibar{}^{\rmc}\gamma_{5}\slashed{\partial}\psi=\psibar{}^{\rmc}\overleftarrow{\slashed{\partial}}\gamma_{5}\psi=0,
\end{eqnarray}
where $k_{\mu}\equiv\psibar\gamma_{5}\gamma_{\mu}\psi$, the axial (pseudo) current density and $\psibar{}^{\rmc}\equiv\psi^{\rmT}C$. These consistency conditions constitute the conservation of current and partial conservation of axial current, as well as a complex consistency condition originally derived by Eliezer \cite{Eliezer-1958}.

\section{Non-abelian SU(2) case}
The $SU(2)$ gauge covariant Dirac equation for a doublet spinor $\Psi$ is
\begin{equation}\label{SU(2) Dirac eqn}
[\rmi\slashed{\partial}-(g/2)\btau\cdot\slashed{\bW}-m]\Psi=0
\end{equation}
where $\tau_{a}$ ($a=1,2,3$) are the non-commutative generators of infinitesimal rotations in doublet space and $W_{a\mu}$ are the Yang-Mills fields, the $SU(2)$ gauge fields analogous to $A_{\mu}$. Explicitly, $\tau_{a}$ are the Pauli matrices, which obey the commutation relations
\begin{equation}
[\tau_{a}/2,\tau_{b}/2]=\rmi\epsilon_{ab}{}^{c}\tau_{c}/2
\end{equation}
where $\epsilon_{abc}$ is the rank-3 Levi-Civita tensor, and is antisymmetric under exchange of any two indices. Note that throughout this paper, we use the Einstein summation convention for the Pauli indices, where a repeated index implies summation, with raising/lowering used to highlight the fact. Now, to derive the charge conjugate of this equation, we must follow the same process involved in charge conjugating the abelian Dirac equation. The goal in the abelian case was to flip the sign of the charge relative to all the other terms, and was achieved by complex conjugating the entire equation, then multiplying it by an invertible matrix $U$, such that $U\gamma_{\mu}^{*}U^{-1}=-\gamma_{\mu}$ and $U\psi^{*}=\psi^{\rmc}$. We follow the same process in the $SU(2)$ case, but not with the presupposition that the sign of $g$ will necessarily be flipped. Complex conjugating and rearranging (\ref{SU(2) Dirac eqn}),
\begin{equation}\label{SU(2) Dirac eqn after complex conjugation}
\{[\rmi\partial_{\nu}+(g/2)\btau^{\rmT}\cdot\bW_{\nu}]\gamma^{\nu*}+m\}\Psi^{*}=0.
\end{equation}
We have used the fact that $\btau$ is Hermitian ($\tau_{a}^{*}=\tau_{a}^{T}$). Multiply (\ref{SU(2) Dirac eqn after complex conjugation}) by $I\otimes U$, with $I$ being the $2\times 2$ identity in doublet space, indicating that $U$ commutes with $\btau$ and acts only on the Dirac spinor degree of freedom
\begin{eqnarray}
&0=\{[\rmi\partial_{\nu}+(g/2)\btau^{\rmT}\cdot\bW_{\nu}]U\gamma^{\nu*}U^{-1}+m\}U\Psi^{*} \nonumber \\
\Rightarrow\ \ &0=\{[\rmi\partial_{\nu}+(g/2)\btau^{\rmT}\cdot\bW_{\nu}]\gamma^{\nu}-m\}\Psi^{\rmc}. \label{SU(2) Dirac charge conjugate equation}
\end{eqnarray}
To make the form of the gauge potential covariant, another step is required which is not present in the abelian case, to convert the $\btau^{\rmT}$ back to $\btau$. This is done by multiplying (\ref{SU(2) Dirac charge conjugate equation}) by $\epsilon\equiv\rmi\tau_{2}$, such that we make use of the Pauli identity $\tau_{a}=-\epsilon\tau_{a}^{\rmT}\epsilon^{-1}$,
\begin{eqnarray}
&0=\{[\rmi\partial_{\nu}+(g/2)\epsilon\btau^{\rmT}\epsilon^{-1}\cdot\bW_{\nu}]\gamma^{\nu}-m\}\epsilon\Psi^{\rmc} \nonumber \\
\Rightarrow\ \ \ &0=[\rmi\slashed{\partial}-(g/2)\btau\cdot\slashed{\bW}-m]\Psi^{\rmic} \label{SU(2) isocharge conjugate Dirac equation}
\end{eqnarray}
where we have defined $\epsilon\Psi^{\rmc}\equiv\Psi^{\rmic}$ as the isospin-charge conjugate (henceforth, IC) spinor. Note that the sign of the coupling constant $g$ has reverted back, so that (\ref{SU(2) isocharge conjugate Dirac equation}) is of exactly the same form as (\ref{SU(2) Dirac eqn}). Mimicking the abelian case (\ref{Emag Dirac eqn compressed}), we rearrange (\ref{SU(2) Dirac eqn}) and (\ref{SU(2) isocharge conjugate Dirac equation}) into the more convenient forms
\begin{eqnarray}
\gamma^{\nu}\btau\cdot\bW_{\nu}\Psi\equiv\tau^{b}\gamma^{\nu}\Psi W_{b\nu}=\Phi \label{SU(2) Dirac eqn compressed} \\
\gamma^{\nu}\btau\cdot\bW_{\nu}\Psi^{\rmic}\equiv\tau^{b}\gamma^{\nu}\Psi^{\rmic}W_{b\nu}=\Phi^{\rmic} \label{SU(2) isocharge conjugate Dirac eqn compressed}
\end{eqnarray}
where $\Phi\equiv2g^{-1}(\rmi\slashed{\partial}-m)\Psi$, $\Phi^{\rmic}\equiv2g^{-1}(\rmi\slashed{\partial}-m)\Psi^{\rmic}$ and we have replaced the triplet vector dot-product notion with a sum over Pauli components $b$. Multiplying equation (\ref{SU(2) Dirac eqn compressed}) by $\Psibar\tau_{a}\gamma_{\mu}$, applying the Dirac identity that produced (\ref{U(1) Dirac eqn after pre-mult}), as well as the Pauli identity
\begin{equation}
\tau_{a}\tau^{b}W_{b\mu}=(\delta_{a}{}^{b}+\rmi\epsilon_{a}{}^{bc}\tau_{c})W_{b\mu}=W_{a\mu}+\rmi\epsilon_{a}{}^{bc}\tau_{c}W_{b\mu}
\end{equation}
we obtain the expression
\begin{eqnarray}\label{SU(2) Dirac eqn after Psi-tau-gamma pre-mult}
\fl\Psibar\Psi W_{a\mu}-\rmi\Psibar\sigma_{\mu}{}^{\nu}\Psi W_{a\nu}+\rmi\epsilon_{a}{}^{bc}\Psibar\tau_{c}\Psi W_{b\mu}+\epsilon_{a}{}^{bc}\Psibar\tau_{c}\sigma_{\mu}{}^{\nu}\Psi W_{b\nu}=\Psibar\tau_{a}\gamma_{\mu}\Phi
\end{eqnarray}
with the form of the IC equation being exactly the same. This is similar to (\ref{U(1) Dirac eqn after pre-mult}), but with ``extra'' terms contracted with the Levi-Civita symbol on the left-hand side. We require a non-abelian analogue of the bispinor relationship (\ref{U(1) bispinor relation}), which is
\begin{equation}\label{SU(2) bispinor relation}
\Psibar{}^{\rmic}(\tau_{i}\otimes\Gamma)\Psi^{\rmic}=-\Psibar(\epsilon^{-1}\tau_{i}{}^{\rmT}\epsilon)\otimes(C^{-1}\Gamma^{\rmT}C)\Psi
\end{equation}
with $\Gamma$ being an element of the Dirac algebra as before, and $\tau_{i}$ an element of the Pauli algebra, where $i=0,1,2,3$ and $i=0$ corresponds to the $2\times2$ identity. Explicit sign relations for particular values of $(\tau_{i}\otimes\Gamma)$ are not given here, although they can easily be calculated following the same method as in (\ref{U(1) bispinor relation}). Following the same method as with the abelian case, we \textit{subtract} the IC of (\ref{SU(2) Dirac eqn after Psi-tau-gamma pre-mult}) from (\ref{SU(2) Dirac eqn after Psi-tau-gamma pre-mult}), apply the appropriate current sign relationships from (\ref{SU(2) bispinor relation}), then rearrange to get
\begin{equation}\label{SU(2) potential before Fierz}
\fl\Psibar\Psi gW_{a\mu}+\epsilon_{a}{}^{bc}\Psibar\tau_{c}\sigma_{\mu}{}^{\nu}\Psi gW_{b\nu}=\rmi(\Psibar\tau_{a}\gamma_{\mu}\slashed{\partial}\Psi-\Psibar\overleftarrow{\slashed{\partial}}\tau_{a}\gamma_{\mu}\Psi)-2m\Psibar\tau_{a}\gamma_{\mu}\Psi.
\end{equation}
We will now define a suite of non-abelian currents, that will be used throughout the rest of this paper:
\numparts
\begin{eqnarray}
J_{i}=\Psibar\tau_{i}\Psi \\
J_{i\mu}=\Psibar\tau_{i}\gamma_{\mu}\Psi \\
S_{i\mu\nu}=\Psibar\tau_{i}\sigma_{\mu\nu}\Psi \\
\Sdual_{i\mu\nu}=\Psibar\tau_{i}\gamma_{5}\sigma_{\mu\nu}\Psi \\
K_{i\mu}=\Psibar\tau_{i}\gamma_{5}\gamma_{\mu}\Psi \\
K_{i}=\Psibar\tau_{i}\gamma_{5}\Psi.
\end{eqnarray}
\endnumparts
There are 64 current densities altogether, excluding $\Sdual_{i\mu\nu}$ from the count, since
\begin{equation}\label{S-dual definition}
\Sdual_{i\mu\nu}=(\rmi/2)\epsilon_{\mu\nu\sigma\rho}S_{i}{}^{\sigma\rho}.
\end{equation}
However, a consequence of the results presented in the next section is that the number of linearly independent current densities is lower than 64, as the $S_{i\mu\nu}$ terms in any expression can be eliminated entirely. Equation (\ref{SU(2) potential before Fierz}) can now be rewritten in the more compact form
\begin{equation}\label{J Dirac eqn invertible}
\left(J_{0}\delta_{\mu}{}^{\nu}\delta_{a}{}^{b}-\epsilon_{a}{}^{cb}S_{c\mu}{}^{\nu}\right)gW_{b\nu}=\rmi(\Psibar\tau_{a}\gamma_{\mu}\slashed{\partial}\Psi-\Psibar\overleftarrow{\slashed{\partial}}\tau_{a}\gamma_{\mu}\Psi)-2mJ_{a\mu}.
\end{equation}
It is apparent that there is an additional matrix term on the left-hand side of (\ref{SU(2) potential before Fierz}) that is not present in the abelian case, which prevents us from immediately finding an inverted form for $W_{a\mu}$. Now, let us consider, as with the abelian case, an alternative formulation of the above expression, by pre-multiplying (\ref{SU(2) Dirac eqn compressed}) and (\ref{SU(2) isocharge conjugate Dirac eqn compressed}) by $\Psibar\tau_{a}\gamma_{5}\gamma_{\mu}$. Applying the sign relationships via (\ref{SU(2) bispinor relation}) and \textit{subtracting} the IC equation from the non-IC equation gives
\begin{equation}\label{K Dirac eqn invertible}
\left(K_{0}\delta_{\mu}{}^{\nu}\delta_{a}{}^{b}-\epsilon_{a}{}^{cb}\Sdual_{c\mu}{}^{\nu}\right)gW_{b\nu}=\rmi(\Psibar\tau_{a}\gamma_{5}\gamma_{\mu}\slashed{\partial}\Psi+\Psibar\overleftarrow{\slashed{\partial}}\tau_{a}\gamma_{5}\gamma_{\mu}\Psi),
\end{equation}
in which the mass term vanishes, as in the abelian case. We could go a step further and \textit{add} the two equations (\ref{J Dirac eqn invertible}) and (\ref{K Dirac eqn invertible}), then divide by the scalar terms to give
\begin{eqnarray}\label{Invertible vector potential equation}
\fl\left[\delta_{\mu}{}^{\nu}\delta_{a}{}^{b}-\frac{\epsilon_{a}{}^{cb}(S_{c\mu}{}^{\nu}+\Sdual_{c\mu}{}^{\nu})}{J_{0}+K_{0}}\right]W_{b\nu} \nonumber \\
=\frac{1}{g}\frac{\rmi(\Psibar\tau_{a}\gamma_{\mu}\slashed{\partial}\Psi+\Psibar\tau_{a}\gamma_{5}\gamma_{\mu}\slashed{\partial}\Psi-\Psibar\overleftarrow{\slashed{\partial}}\tau_{a}\gamma_{\mu}\Psi+\Psibar\overleftarrow{\slashed{\partial}}\tau_{a}\gamma_{5}\gamma_{\mu}\Psi)-2mJ_{a\mu}}{J_{0}+K_{0}}.
\end{eqnarray}
The matrix on the left-hand side is of the form $(I-N)$, which is invertible by way of the Neumann series
\begin{equation}\label{Neumann series definition}
(I-N)^{-1}=\sum^{\infty}_{n=0}N^{n}=I+N+N^{2}+...
\end{equation}
for $N\in\mathbb{C}^{n\times n}$, which has condition of convergence $\rho(N)<1$, where $\rho(N)$ is the spectral radius of $N$ \cite{Hogben-2007}. In the following section, we will show that the terms in the expansion of $(I-N)^{-1}$ can be converted from contractions involving spin-current tensors to contractions involving $J, K$ Lorentz scalar and vector currents exclusively, by deriving appropriate Fierz identities. Firstly, a brief comment on some of the various other objects which the Dirac and IC Dirac equations, (\ref{SU(2) Dirac eqn compressed}) and (\ref{SU(2) isocharge conjugate Dirac eqn compressed}), can be multiplied by to form bispinor-field coupling expressions. We could multiply either of these equations on the left by objects of the form $\Psibar\tau_{i}\Gamma$, where $\Gamma$ is an irreducible element of the Dirac basis (including the dual of the rank-2 tensor, $\gamma_{5}\sigma_{\mu\nu}$), then either add or subtract the resulting equations. For example, consider the case where $\Gamma=I$, $i=0$, then we multiply (\ref{SU(2) Dirac eqn compressed}) and (\ref{SU(2) isocharge conjugate Dirac eqn compressed}) by $\Psibar$ and $\Psibar{}^{\rmic}$ respectively. Subtracting the IC equation from the non-IC equation gives
\begin{equation}
(g/2)J^{b\nu}W_{b\nu}=(\rmi/2)(\Psibar\spart\Psi-\Psibar\lspart\Psi)-mJ_{0},
\end{equation}
and adding them gives
\begin{equation}
\Psibar\lspart\Psi=-\Psibar\spart\Psi.
\end{equation}
We could choose to eliminate the bispinor with the left-acting derivative operator by substituting the second equation into the first, resulting in the expression
\begin{equation}
(g/2)J^{b\nu}W_{b\nu}=\rmi\Psibar\spart\Psi-mJ_{0}.
\end{equation}
This equation describes the coupling of the Lorentz vector current $J_{a\mu}$ with the vector potential field via contraction of both the Pauli and Lorentz indices, resulting in a sum of Lorentz and Pauli scalar terms. Unlike the invertible cases discussed above, this equation has no free indices and therefore can not be inverted via multiplication by an appropriate matrix. Since these types of equations may provide valuable information in future studies, all $\Psibar\tau_{i}\Gamma$ multiplication options are listed in appendix B.

\section{Non-abelian Fierz identities}
Consider the $8\times8$ matrix formed by the product $\Psi\Psibar$. In the $SU(2)$ doublet degree of freedom, this is a $2\times2$ matrix, and in the Dirac spinor degree of freedom, it is $4\times4$ matrix. In the pure Dirac case, the product of two Dirac spinors can be expanded via a Fierz expansion
\begin{eqnarray}\label{Dirac Fierz}
\fl\psi\psibar=\sum_{R=1}^{16}a_{R}\Gamma_{R}=&(1/4)(\psibar\psi)I+(1/4)(\psibar\gamma_{\mu}\psi)\gamma^{\mu}+(1/8)(\psibar\sigma_{\mu\nu}\psi)\sigma^{\mu\nu} \nonumber \\
&-(1/4)(\psibar\gamma_{5}\gamma_{\mu}\psi)\gamma_{5}\gamma^{\mu}+(1/4)(\psibar\gamma_{5}\psi)\gamma_{5}.
\end{eqnarray}
There are 16 terms in the sum, and the coefficients $a_{R}$ are the Dirac bispinors $\psibar\Gamma_{R}\psi$, multiplied by a numerical constant, with $\Gamma_{R}$ being the $R$th element of the sixteen-component basis of the Dirac algebra. It is interesting to note that extensions of Fierz expansions such as this have been described in arbitrary higher dimensions in \cite{Delbourgo-Prasad-1974}. Now, in the pure $2\times2$ case, the Fierz expansion for the matrix formed by the product of two doublet spinors $v$ is
\begin{equation}
vv^{\dagger}=\sum_{i=0}^{3}c_{i}\tau_{i}=(1/2)(v^{\dagger}v)\tau_{0}+(1/2)(v^{\dagger}\tau_{a}v)\tau^{a}
\end{equation}
with $a=1,2,3$, to make four terms in the sum in total. As discussed before, $\tau_{a}$ are the Pauli matrices, and $\tau_{0}$ is the $2\times2$ identity. Coefficients $c_{i}$ are pure $SU(2)$ isospin bispinors $v^{\dagger}\tau_{i}v$. Now, the basis of the $\Psi\Psibar$ Fierz expansion is the tensor product of the Dirac and Pauli bases
\begin{eqnarray}\label{Pauli-Dirac Fierz expansion}
\fl\Psi\Psibar=\sum_{i=0}^{3}\sum_{R=1}^{16}a_{Ri}(\Gamma_{R}\otimes\tau_{i})=(1/8)J_{i}(I\otimes\tau^{i})+(1/8)J_{i\mu}(\gamma^{\mu}\otimes\tau^{i}) \nonumber \\
+(1/16)S_{i\mu\nu}(\sigma^{\mu\nu}\otimes\tau^{i})-(1/8)K_{i\mu}(\gamma_{5}\gamma^{\mu}\otimes\tau^{i})+(1/8)K_{i}(\gamma_{5}\otimes\tau^{i})
\end{eqnarray}
where the coefficients are the $SU(2)$ bispinors as previously defined, and are derived by pre-multiplying (\ref{Pauli-Dirac Fierz expansion}) by an element of $(\Gamma_{R}\otimes\tau_{i})$, utilizing trace identities, and then solving for the leftover $a_{Ri}$. Henceforth, we will exclude the tensor product symbol explicitly, however its presence is implied in any product of Dirac and Pauli matrices.

The Fierz expansion can used to expand products of non-abelian currents in terms of other currents in the Dirac-Pauli algebra, for example
\begin{eqnarray}
\fl J_{a}{}^{\mu}K_{b}{}^{\nu}=\Psibar\tau_{a}\gamma^{\mu}(\Psi\Psibar)\tau_{b}\gamma_{5}\gamma^{\nu}\Psi \nonumber \\
=-(1/8)J_{i}\Psibar\gamma_{5}\gamma^{\mu}\gamma^{\nu}\tau_{a}\tau^{i}\tau_{b}\Psi+(1/8)J_{i\sigma}\Psibar\gamma_{5}\gamma^{\mu}\gamma^{\sigma}\gamma^{\nu}\tau_{a}\tau^{i}\tau_{b}\Psi \nonumber \\
\qquad -(1/16)S_{i\sigma\epsilon}\Psibar\gamma_{5}\gamma^{\mu}\sigma^{\sigma\epsilon}\gamma^{\nu}\tau_{a}\tau^{i}\tau_{b}\Psi \nonumber \\
\qquad +(1/8)K_{i\sigma}\Psibar\gamma^{\mu}\gamma^{\sigma}\gamma^{\nu}\tau_{a}\tau^{i}\tau_{b}\Psi+(1/8)K_{i}\Psibar\gamma^{\mu}\gamma^{\nu}\tau_{a}\tau^{i}\tau_{b}\Psi
\end{eqnarray}
which, after a lengthy expansion, converting all of the Dirac and Pauli matrix products to sums of irreducible terms gives
\begin{eqnarray}
\fl J_{a}{}^{\mu}K_{b}{}^{\nu}=(1/4)[\rmi J_{a}\Sdual_{b}{}^{\mu\nu}+\rmi J_{b}\Sdual_{a}{}^{\mu\nu}-\rmi K_{a}S_{b}{}^{\mu\nu}-\rmi K_{b}S_{a}{}^{\mu\nu}+J_{a}{}^{\mu}K_{b}{}^{\nu} \nonumber \\
+J_{a}{}^{\nu}K_{b}{}^{\mu}+J_{b}{}^{\mu}K_{a}{}^{\nu}+J_{b}{}^{\nu}K_{a}{}^{\mu}-J_{a}{}_{\sigma}K_{b}{}^{\sigma}\eta^{\mu\nu}-J_{b}{}_{\sigma}K_{a}{}^{\sigma}\eta^{\mu\nu} \nonumber \\
+\delta_{ab}(\rmi J_{0}\Sdual_{0}{}^{\mu\nu}-\rmi J_{c}\Sdual^{c\mu\nu}-\rmi K_{0}S_{0}{}^{\mu\nu}+\rmi K_{c}S^{c\mu\nu}+J_{0}{}^{\mu}K_{0}{}^{\nu} \nonumber \\
+J_{0}{}^{\nu}K_{0}{}^{\mu}-J_{c}{}^{\mu}K^{c\nu}-J_{c}{}^{\nu}K^{c\mu}-J_{0\sigma}K_{0}{}^{\sigma}\eta^{\mu\nu}+J_{c\sigma}K^{c\sigma}\eta^{\mu\nu})] \nonumber \\
+(1/4)\epsilon_{ab}{}^{c}[-\rmi J_{0}K_{c}\eta^{\mu\nu}+\rmi K_{0}J_{c}\eta^{\mu\nu}+J_{c\sigma}J_{0\lambda}\epsilon^{\mu\nu\sigma\lambda}+K_{c\sigma}K_{0\lambda}\epsilon^{\mu\nu\sigma\lambda} \nonumber \\
+(1/2)\rmi(-S_{0}{}^{\mu}{}_{\sigma}\Sdual_{c}{}^{\sigma\nu}-S_{0}{}^{\nu}{}_{\sigma}\Sdual_{c}{}^{\sigma\mu}+S_{c}{}^{\mu}{}_{\sigma}\Sdual_{0}{}^{\sigma\nu}+S_{c}{}^{\nu}{}_{\sigma}\Sdual_{0}{}^{\sigma\mu})].
\end{eqnarray}
In the set of Fierz identities for $JK$ vector products, we call this the $a-b$ case. There are three other cases of vector current products: $J_{a}{}^{\mu}K_{0}{}^{\nu}$, $J_{0}{}^{\mu}K_{a}{}^{\nu}$ and $J_{0}{}^{\mu}K_{0}{}^{\nu}$, which we call the $a-0$, $0-a$ and $0-0$ cases respectively. Due to their long-winded nature, we leave the derivation and listing of these Fierz identities to appendix C. Now, if we wish to express the inverted form of the vector potential equation (\ref{Invertible vector potential equation}) independently of the rank-2 spin current (skew) tensor $S_{i}{}^{\mu\nu}$ and its dual, we must obtain an identity to describe $S_{i}{}^{\mu\nu}$ solely in terms of $J$, $K$ scalar and vector current densities. Since $S_{i}{}^{\mu\nu}$ is antisymmetric under exchange of its two Lorentz indices, we should form antisymmetric terms from the two-vector product $J_{a}{}^{\mu}K_{b}{}^{\nu}$, then Fierz expand using (\ref{Pauli-Dirac Fierz expansion}) and solve for $S_{0}{}^{\mu\nu}$, $S_{a}{}^{\mu\nu}$, $\Sdual_{0}{}^{\mu\nu}$ or $\Sdual_{a}{}^{\mu\nu}$.

Let us define the suite of 16 \textit{abelian} currents as follows
\numparts
\begin{eqnarray}
\sigma=\psibar\psi \\
j_{\mu}=\psibar\gamma_{\mu}\psi \\
s_{\mu\nu}=\psibar\sigma_{\mu\nu}\psi \\
\sdual_{\mu\nu}=\psibar\gamma_{5}\sigma_{\mu\nu}\psi \\
k_{\mu}=\psibar\gamma_{5}\gamma_{\mu}\psi \\
\omega=\psibar\gamma_{5}\psi
\end{eqnarray}
\endnumparts
where the dual of the rank-2 skew tensor current may be calculated as in (\ref{S-dual definition}). If we consider the Fierz identity for the abelian version of the rank-2 spin current tensor \cite{Crawford-1985}, \cite{Kaempffer-1981},
\begin{equation}\label{S abelian Fierz identity}
s_{\mu\nu}=\frac{[\sigma\epsilon^{\mu\nu\rho\kappa}-\rmi\omega(\delta_{\mu}{}^{\rho}\delta_{\nu}{}^{\kappa}-\delta_{\mu}{}^{\kappa}\delta_{\nu}{}^{\rho})]j_{\rho}k_{\kappa}}{\sigma^{2}-\omega^{2}}
\end{equation}
we can see that, in the non-abelian case, we should consider Fierz expansions of antisymmetric current combinations with Lorentz structure of the form $J^{\mu}K^{\nu}-J^{\nu}K^{\mu}$ and $\epsilon^{\mu\nu\rho\kappa}J_{\rho}K_{\kappa}$. Due to the presence of the extra internal Pauli index, we need to take consideration of how this will vary the form of $S_{i}{}^{\mu\nu}$ compared with $s^{\mu\nu}$. As discussed in appendix D, the correct approach is to treat the derivation of $S_{0}{}^{\mu\nu}$ and $S_{a}{}^{\mu\nu}$ separately. For the $i=0$ case, we calculate the Fierz identities for $J_{0}{}^{\mu}K_{0}{}^{\nu}-J_{a}{}^{\nu}K^{a\mu}$, and $J_{a}{}^{\mu}K^{a\nu}-J_{0}{}^{\nu}K_{0}{}^{\mu}$, then add to form
\begin{equation}
J_{i}{}^{\mu}K^{i\nu}-J_{i}{}^{\nu}K^{i\mu}=2\rmi(J_{0}\Sdual_{0}{}^{\mu\nu}-K_{0}S_{0}{}^{\mu\nu}).
\end{equation}
The antisymmetric part is calculated by adding $\epsilon^{\mu\nu\rho\kappa}J_{0\rho}K_{0\kappa}$ and $\epsilon^{\mu\nu\rho\kappa}J_{a\rho}K^{a}{}_{\kappa}$ to form
\begin{equation}
\epsilon^{\mu\nu\rho\kappa}J_{i\rho}K^{i}{}_{\kappa}=2(J_{0}S_{0}{}^{\mu\nu}-K_{0}\Sdual_{0}{}^{\mu\nu}).
\end{equation}
Taking the combination
\begin{eqnarray}\label{Antisymmetric combination for S0}
\fl J_{0}\epsilon^{\mu\nu\rho\kappa}J_{i\rho}K^{i}{}_{\kappa}-\rmi K_{0}(J_{i}{}^{\mu}K^{i\nu}-J_{i}{}^{\nu}K^{i\mu}) \nonumber \\
=2(J_{0}^{2}S_{0}{}^{\mu\nu}-J_{0}K_{0}\Sdual_{0}{}^{\mu\nu}+J_{0}K_{0}\Sdual_{0}{}^{\mu\nu}-K_{0}^{2}S_{0}{}^{\mu\nu}),
\end{eqnarray}
the middle two terms on the right-hand side cancel, and we can rearrange to obtain the expression
\begin{equation}\label{S0 Fierz identity}
S_{0}{}^{\mu\nu}=(1/2)(J_{0}^{2}-K_{0}^{2})^{-1}[J_{0}\epsilon^{\mu\nu}{}_{\rho\kappa}-\rmi K_{0}(\delta_{\rho}{}^{\mu}\delta_{\kappa}{}^{\nu}-\delta_{\rho}{}^{\nu}\delta_{\kappa}{}^{\mu})]J_{i}{}^{\rho}K^{i\kappa}.
\end{equation}
The abelian (\ref{S abelian Fierz identity}) and non-abelian (\ref{S0 Fierz identity}) cases share a close similarity, with the main differences being the factor of $1/2$, and the sum over the internal Pauli index in the non-abelian case. Similarly, we can switch the terms that $J_{0}$ and $K_{0}$ multiply in (\ref{Antisymmetric combination for S0}) to obtain an expression for the dual
\begin{equation}\label{Sdual0 Fierz identity}
\Sdual_{0}{}^{\mu\nu}=(1/2)(J_{0}^{2}-K_{0}^{2})^{-1}[K_{0}\epsilon^{\mu\nu}{}_{\rho\kappa}-\rmi J_{0}(\delta_{\rho}{}^{\mu}\delta_{\kappa}{}^{\nu}-\delta_{\rho}{}^{\nu}\delta_{\kappa}{}^{\mu})]J_{i}{}^{\rho}K^{i\kappa}.
\end{equation}
Note that we can check the validity of this dual identity by using the defining identity (\ref{S-dual definition}) on either (\ref{S0 Fierz identity}) or (\ref{Sdual0 Fierz identity}). Now taking a step further, the scalar currents inside the square brackets can be eliminated by taking the sum or difference of (\ref{S0 Fierz identity}) and (\ref{Sdual0 Fierz identity})
\begin{equation}
S_{0}{}^{\mu\nu}\pm\Sdual_{0}{}^{\mu\nu}=(1/2)(J_{0}\mp K_{0})^{-1}[\epsilon^{\mu\nu}{}_{\rho\kappa}\mp\rmi(\delta_{\rho}{}^{\mu}\delta_{\kappa}{}^{\nu}-\delta_{\rho}{}^{\nu}\delta_{\kappa}{}^{\mu})]J_{i}{}^{\rho}K^{i\kappa}.
\end{equation}
Now, in order to form an expression for $S_{a}{}^{\mu\nu}$, we again need to consider antisymmetric combinations of $JK$ vector current products, but in such a way as to have the Pauli vector triplet index $i=a=1,2,3$ present in first order only, combined with the Pauli scalar singlet index $i=0$. That is, we shall be dealing with the \textit{rank-1} $JK$ Pauli vector combinations, the $0-a$ and $a-0$ cases, as opposed to the $0-0$  \textit{rank-0} or $a-b$ \textit{rank-2} Pauli index cases. As discussed in appendix D, the appropriate Lorentz antisymmetric combinations are
\begin{eqnarray}
\fl(J_{a}{}^{\mu}K_{0}{}^{\nu}+J_{0}{}^{\mu}K_{a}{}^{\nu})-(J_{a}{}^{\nu}K_{0}{}^{\mu}+J_{0}{}^{\nu}K_{a}{}^{\mu}) \nonumber \\
=\rmi J_{0}\Sdual_{a}{}^{\mu\nu}+\rmi J_{a}\Sdual_{0}{}^{\mu\nu}-\rmi K_{0}S_{a}{}^{\mu\nu}-\rmi K_{a}S_{0}{}^{\mu\nu}
\end{eqnarray}
as well as
\begin{eqnarray}
\epsilon^{\mu\nu\rho\kappa}(J_{a\rho}K_{0\kappa}+J_{0\rho}K_{a\kappa})=J_{0}S_{a}{}^{\mu\nu}+J_{a}S_{0}{}^{\mu\nu}-K_{0}\Sdual_{a}{}^{\mu\nu}-K_{a}\Sdual_{0}{}^{\mu\nu}.
\end{eqnarray}
Taking the combination
\begin{eqnarray}\label{Antisymmetric combination for Sa}
\fl J_{0}\epsilon^{\mu\nu\rho\kappa}(J_{a\rho}K_{0\kappa}+J_{0\rho}K_{a\kappa})-\rmi K_{0}[(J_{a}{}^{\mu}K_{0}{}^{\nu}+J_{0}{}^{\mu}K_{a}{}^{\nu})-(J_{a}{}^{\nu}K_{0}{}^{\mu}+J_{0}{}^{\nu}K_{a}{}^{\mu})] \nonumber \\
=(J_{0}^{2}-K_{0}^{2})S_{a}{}^{\mu\nu}+(J_{0}J_{a}-K_{0}K_{a})S_{0}{}^{\mu\nu}+(K_{0}J_{a}-J_{0}K_{a})\Sdual_{0}{}^{\mu\nu}
\end{eqnarray}
we can see that after rearranging to solve for $S_{a}{}^{\mu\nu}$, we need to substitute the identities (\ref{S0 Fierz identity}) and (\ref{Sdual0 Fierz identity}) to eliminate the Pauli singlet $(i=0)$ skew tensor current densities. The final expression is
\begin{eqnarray}\label{Sa Fierz identity}
\fl S_{a}{}^{\mu\nu}=(J_{0}^{2}-K_{0}^{2})^{-1}[J_{0}\epsilon^{\mu\nu}{}_{\rho\kappa}-\rmi K_{0}(\delta_{\rho}{}^{\mu}\delta_{\kappa}{}^{\nu}-\delta_{\rho}{}^{\nu}\delta_{\kappa}{}^{\mu})](J_{a}{}^{\rho}K_{0}{}^{\kappa}+J_{0}{}^{\rho}K_{a}{}^{\kappa}) \nonumber \\
-\frac{J_{0}^{2}+K_{0}^{2}}{2(J_{0}^{2}-K_{0}^{2})^{2}}[J_{a}\epsilon^{\mu\nu}{}_{\rho\kappa}+\rmi K_{a}(\delta_{\rho}{}^{\mu}\delta_{\kappa}{}^{\nu}-\delta_{\rho}{}^{\nu}\delta_{\kappa}{}^{\mu})]J_{i}{}^{\rho}K^{i\kappa} \nonumber \\
+\frac{J_{0}K_{0}}{(J_{0}^{2}-K_{0}^{2})^{2}}[K_{a}\epsilon^{\mu\nu}{}_{\rho\kappa}+\rmi J_{a}(\delta_{\rho}{}^{\mu}\delta_{\kappa}{}^{\nu}-\delta_{\rho}{}^{\nu}\delta_{\kappa}{}^{\mu})]J_{i}{}^{\rho}K^{i\kappa},
\end{eqnarray}
which bears less resemblance to the abelain case (\ref{S abelian Fierz identity}) than the Pauli singlet case (\ref{S0 Fierz identity}) does. To calculate the dual, we follow exactly the same process, but switch the $J_{0}$ and $K_{0}$ in (\ref{Antisymmetric combination for Sa}), which after some rearrangement and substitution gives
\begin{eqnarray}\label{Sduala Fierz identity}
\fl\Sdual_{a}{}^{\mu\nu}=(J_{0}^{2}-K_{0}^{2})^{-1}[K_{0}\epsilon^{\mu\nu}{}_{\rho\kappa}-\rmi J_{0}(\delta_{\rho}{}^{\mu}\delta_{\kappa}{}^{\nu}-\delta_{\rho}{}^{\nu}\delta_{\kappa}{}^{\mu})](J_{a}{}^{\rho}K_{0}{}^{\kappa}+J_{0}{}^{\rho}K_{a}{}^{\kappa}) \nonumber \\
+\frac{J_{0}^{2}+K_{0}^{2}}{2(J_{0}^{2}-K_{0}^{2})^{2}}[K_{a}\epsilon^{\mu\nu}{}_{\rho\kappa}+\rmi J_{a}(\delta_{\rho}{}^{\mu}\delta_{\kappa}{}^{\nu}-\delta_{\rho}{}^{\nu}\delta_{\kappa}{}^{\mu})]J_{i}{}^{\rho}K^{i\kappa} \nonumber \\
-\frac{J_{0}K_{0}}{(J_{0}^{2}-K_{0}^{2})^{2}}[J_{a}\epsilon^{\mu\nu}{}_{\rho\kappa}+\rmi K_{a}(\delta_{\rho}{}^{\mu}\delta_{\kappa}{}^{\nu}-\delta_{\rho}{}^{\nu}\delta_{\kappa}{}^{\mu})]J_{i}{}^{\rho}K^{i\kappa}.
\end{eqnarray}
Again, this can be confirmed by using (\ref{S-dual definition}). As in the $i=0$ case, taking the sum or difference of these two identities removes the current terms from inside the square brackets. Following some straightforward algebraic manipulation, we obtain a somewhat simpler form
\begin{eqnarray}\label{Sa Sduala sum/difference}
\fl S_{a}{}^{\mu\nu}\pm\Sdual_{a}{}^{\mu\nu}=(1/2)(J_{0}\mp K_{0})^{-2}[\epsilon^{\mu\nu}{}_{\rho\kappa}\mp\rmi(\delta_{\rho}{}^{\mu}\delta_{\kappa}{}^{\nu}-\delta_{\rho}{}^{\nu}\delta_{\kappa}{}^{\mu})] \nonumber \\
\cdot[2(J_{0}\mp K_{0})(J_{a}{}^{\rho}K_{0}{}^{\kappa}+J_{0}{}^{\rho}K_{a}{}^{\kappa})-(J_{a}\mp K_{a})J_{i}{}^{\rho}K^{i\kappa}].
\end{eqnarray}
The relative simplicity of (\ref{Sa Sduala sum/difference}) compared with (\ref{Sa Fierz identity}) and (\ref{Sduala Fierz identity}) is the reason we chose to pursue an invertible Dirac equation of the form (\ref{Invertible vector potential equation}) as opposed to (\ref{J Dirac eqn invertible}) or (\ref{K Dirac eqn invertible}) alone. In particular, the extra simplicity will have a profound impact on the complexity of the higher-power terms in the Neumann series form of the inverse matrix.

\section{Conclusions}
In this study, the inversion of the Dirac equation to solve for the vector potential has been extended to the non-abelian case, the $SU(2)$ gauge field specifically. An extension of the charge conjugation operation was made in order to make the form of the gauge field generators covariant, and the algebraic system tractable. In analogy to previous studies performed on the abelian case, the non-abelian Dirac system was re-written in terms of bispinor current densities, by way of multiplication by the terms $\Psibar\tau_{a}\gamma_{\mu}$ and $\Psibar\tau_{a}\gamma_{5}\gamma_{\mu}$. Combining these equations, an invertible form was achieved, provided we made use of a Neumann expansion to describe the form of the inverse matrix. In order to eliminate the rank-2 skew tensor current densities from the Neumann expansion, we were motivated to derive appropriate Fierz identities by considering antisymmetric combinations of $J_{i\mu}K_{j\nu}$ current density products. Expressions for $S_{0\mu\nu}$, $\Sdual_{0\mu\nu}$, $S_{a\mu\nu}$ and $\Sdual_{a\mu\nu}$ were subsequently derived, and the convenient linear combinations $S_{0\mu\nu}\pm\Sdual_{0\mu\nu}$ and $S_{a\mu\nu}\pm\Sdual_{a\mu\nu}$ were formed. Some options for further work include deriving consistency conditions, analogous to those presented at the end of section 2, for the $SU(2)$ case, as well as describing explicitly the conditions for the convergence of the inverse matrix Neumann series. A broader study of the non-abelian Fierz identities is also in order, in particular obtaining a complete minimal set, from which all other redundant Fierz identities can be derived. The scope of the inversion may also be extended to $SU(2)\times U(1)$ and $SU(3)$ gauge fields. In the latter case, the lack of an extended analogue of the $SU(2)$ isospin-charge conjugate spinors $\Psi^{\rmic}$ suggests that the analysis of Fierz identities will have to confront a yet more involved set of bispinor current densities, and the Dirac equation, a more complicated inversion calculation.

\appendix
\section{Pauli and Dirac Identities}
Here follows various Pauli and Dirac matrix identities that are used throughout this paper.
\subsection{Pauli Identities}
\begin{eqnarray}
\tau_{a}\tau_{b}=\delta_{ab}+\rmi\epsilon_{abd}\tau^{d}
\end{eqnarray}
\begin{eqnarray}
\tau_{a}\tau_{c}\tau_{b}=\tau_{a}\delta_{bc}+\tau_{b}\delta_{ac}-\tau_{c}\delta_{ab}-\rmi\epsilon_{abc}
\end{eqnarray}

\subsection{Dirac Identities}
\begin{eqnarray}
\fl\gamma^{\mu}\gamma^{\nu}=\eta^{\mu\nu}-\rmi\sigma^{\mu\nu}
\end{eqnarray}
\begin{eqnarray}
\fl\gamma^{\mu}\gamma^{\nu}\gamma^{\lambda}=\eta^{\mu\nu}\gamma^{\lambda}+\eta^{\nu\lambda}\gamma^{\mu}-\eta^{\mu\lambda}\gamma^{\nu}-\rmi\epsilon^{\mu\nu\lambda\sigma}\gamma_{5}\gamma_{\sigma}
\end{eqnarray}
\begin{eqnarray}
\fl\gamma^{\mu}\gamma^{\nu}\gamma^{\sigma}\gamma^{\epsilon}=\eta^{\mu\nu}\eta^{\sigma\epsilon}+\eta^{\nu\sigma}\eta^{\mu\epsilon}-\eta^{\mu\sigma}\eta^{\nu\epsilon}-\rmi\eta^{\mu\nu}\sigma^{\sigma\epsilon}-\rmi\eta^{\nu\sigma}\sigma^{\mu\epsilon} \nonumber \\
+\rmi\eta^{\mu\sigma}\sigma^{\nu\epsilon}+\rmi\eta^{\mu\epsilon}\sigma^{\sigma\nu}+\rmi\eta^{\nu\epsilon}\sigma^{\mu\sigma}+\rmi\eta^{\sigma\epsilon}\sigma^{\nu\mu}-\rmi\epsilon^{\mu\nu\sigma\epsilon}\gamma_{5}
\end{eqnarray}
\begin{eqnarray}
\fl\gamma^{\epsilon}\sigma^{\mu\nu}=\rmi\eta^{\epsilon\mu}\gamma^{\nu}-\rmi\eta^{\epsilon\nu}\gamma^{\mu}+\epsilon^{\mu\nu\epsilon\sigma}\gamma_{5}\gamma_{\sigma}
\end{eqnarray}
\begin{eqnarray}
\fl\sigma^{\mu\nu}\gamma^{\epsilon}=\rmi\eta^{\nu\epsilon}\gamma^{\mu}-\rmi\eta^{\mu\epsilon}\gamma^{\nu}+\epsilon^{\mu\nu\epsilon\sigma}\gamma_{5}\gamma_{\sigma}
\end{eqnarray}
\begin{eqnarray}
\fl\gamma^{\mu}\sigma^{\sigma\epsilon}\gamma^{\nu}=\rmi\eta^{\epsilon\nu}\eta^{\mu\sigma}-\rmi\eta^{\sigma\nu}\eta^{\mu\epsilon}+\eta^{\epsilon\nu}\sigma^{\mu\sigma}-\eta^{\sigma\nu}\sigma^{\mu\epsilon}-\epsilon^{\sigma\epsilon\nu\mu}\gamma_{5}+\rmi\epsilon^{\sigma\epsilon\nu\lambda}\gamma_{5}\sigma^{\mu}{}_{\lambda}
\end{eqnarray}
\begin{eqnarray}
\fl-\epsilon^{\lambda\rho\sigma\epsilon}\epsilon_{\lambda}{}^{\mu\nu\tau}=\eta^{\rho\mu}\eta^{\sigma\nu}\eta^{\epsilon\tau}-\eta^{\rho\mu}\eta^{\epsilon\nu}\eta^{\sigma\tau}+\eta^{\rho\nu}\eta^{\sigma\tau}\eta^{\epsilon\mu}-\eta^{\rho\nu}\eta^{\epsilon\tau}\eta^{\sigma\mu}+\eta^{\rho\tau}\eta^{\sigma\mu}\eta^{\epsilon\nu} \nonumber \\
-\eta^{\rho\tau}\eta^{\epsilon\mu}\eta^{\sigma\nu}
\end{eqnarray}

\section{Current-Field Coupling Identities}
Here we list the system of expressions that result when we multiply the $SU(2)$ gauge covariant Dirac equation and its isospin-charge conjugate equation (IC equation),
\begin{eqnarray}
\tau^{b}\gamma^{\nu}W_{b\nu}\Psi=(2/g)(\rmi\spart-m)\Psi
\end{eqnarray}
\begin{eqnarray}
\tau^{b}\gamma^{\nu}W_{b\nu}\Psi^{\rmic}=(2/g)(\rmi\spart-m)\Psi^{\rmic},
\end{eqnarray}
from the left by a matrix of the general form $\Psibar\tau_{i}\Gamma$. For each of these matrices, we will obtain three equations, the order of the list being: (a) subtract IC from non-IC equation, (b) add IC from non-IC equation, and (c) the result obtained by combining (a) and (b) to eliminate the bispinor with left-acting derivate operator, $\lspart$. Note that the equation we use for substitution will we written in terms of the aforementioned $\lspart$ bispinor.

\vspace{5mm}

\noindent Multiply by $\Psibar$:
\begin{eqnarray}
(g/2)J^{b\nu}W_{b\nu}=(\rmi/2)(\Psibar\spart\Psi-\Psibar\lspart\Psi)-mJ_{0}
\end{eqnarray}
\begin{eqnarray}
\Psibar\lspart\Psi=-\Psibar\spart\Psi
\end{eqnarray}
\begin{eqnarray}
(g/2)J^{b\nu}W_{b\nu}=\rmi\Psibar\spart\Psi-mJ_{0}
\end{eqnarray}
Multiply by $\Psibar\tau_{a}$:
\begin{eqnarray}
\Psibar\lspart\tau_{a}\Psi=g\epsilon_{a}{}^{bc}J_{c}{}^{\nu}W_{b\nu}-\Psibar\tau_{a}\spart\Psi
\end{eqnarray}
\begin{eqnarray}
(g/2)J_{0}{}^{\nu}W_{a\nu}=(\rmi/2)(\Psibar\tau_{a}\spart\Psi-\Psi\lspart\tau_{a}\Psi)-mJ_{a}
\end{eqnarray}
\begin{eqnarray}
(g/2)(J_{0}{}^{\nu}\delta_{a}{}^{b}+\rmi J_{c}{}^{\nu}\epsilon_{a}{}^{bc})W_{b\nu}=\rmi\Psibar\tau_{a}\spart\Psi-mJ_{a}
\end{eqnarray}
Multiply by $\Psibar\gamma_{5}$:
\begin{eqnarray}
\Psibar\lspart\gamma_{5}\Psi=\Psibar\gamma_{5}\spart\Psi+2imK_{0}
\end{eqnarray}
\begin{eqnarray}
(g/2)K^{b\nu}W_{b\nu}=(\rmi/2)(\Psibar\gamma_{5}\spart\Psi+\Psibar\lspart\gamma_{5}\Psi)
\end{eqnarray}
\begin{eqnarray}
(g/2)K^{b\nu}W_{b\nu}=\rmi\Psibar\gamma_{5}\spart\Psi-mK_{0}
\end{eqnarray}
Multiply by $\Psibar\tau_{a}\gamma_{5}$:
\begin{eqnarray}
\Psibar\lspart\gamma_{5}\tau_{a}\Psi=-g\rmi K_{0}{}^{\nu}W_{a\nu}-\Psibar\gamma_{5}\tau_{a}\spart\Psi
\end{eqnarray}
\begin{eqnarray}
\rmi(g/2)\epsilon_{a}{}^{bc}K_{c}{}^{\nu}W_{b\nu}=(\rmi/2)(\Psibar\gamma_{5}\tau_{a}\spart\Psi-\Psibar\lspart\gamma_{5}\tau_{a}\Psi)-mK_{a}
\end{eqnarray}
\begin{eqnarray}
(g/2)(K_{0}{}^{\nu}\delta_{a}{}^{b}+\rmi K_{c}{}^{\nu}\epsilon_{a}{}^{bc})W_{b\nu}=\rmi\Psibar\gamma_{5}\tau_{a}\spart\Psi-mK_{a}
\end{eqnarray}
Multiply by $\Psibar\gamma_{\mu}$:
\begin{eqnarray}
\Psibar\lspart\gamma_{\mu}\Psi=-\Psibar\gamma_{\mu}\spart\Psi-gS^{b}{}_{\mu}{}^{\nu}W_{b\nu}
\end{eqnarray}
\begin{eqnarray}
(g/2)J^{b}W_{b\mu}=(\rmi/2)(\Psibar\gamma_{\mu}\spart\Psi-\Psibar\lspart\gamma_{\mu}\Psi)-mJ_{0\mu}
\end{eqnarray}
\begin{eqnarray}
(g/2)(J^{b}\delta_{\mu}{}^{\nu}-\rmi S^{b}{}_{\mu}{}^{\nu})W_{b\nu}=\rmi\Psibar\gamma_{\mu}\spart\Psi-mJ_{0\mu}
\end{eqnarray}
Multiply by $\Psibar\tau_{a}\gamma_{\mu}$:
\begin{eqnarray}
\fl(g/2)(J_{0}\delta_{\mu}{}^{\nu}\delta_{a}{}^{b}+S_{c\mu}{}^{\nu}\epsilon_{a}{}^{bc})W_{b\nu}=(\rmi/2)(\Psibar\tau_{a}\gamma_{\mu}\spart\Psi-\Psibar\lspart\tau_{a}\gamma_{\mu}\Psi)-mJ_{a\mu}
\end{eqnarray}
\begin{eqnarray}
\fl\Psibar\lspart\tau_{a}\gamma_{\mu}\Psi=g(J_{c}\delta_{\mu}{}^{\nu}\epsilon_{a}{}^{bc}-S_{0\mu}{}^{\nu}\delta_{a}{}^{b})W_{b\mu}-\Psibar\tau_{a}\gamma_{\mu}\spart\Psi
\end{eqnarray}
\begin{eqnarray}
\fl(g/2)(J_{0}\delta_{\mu}{}^{\nu}\delta_{a}{}^{b}+\rmi J_{c}\delta_{\mu}{}^{\nu}\epsilon_{a}{}^{bc}-\rmi S_{0\mu}{}^{\nu}\delta_{a}{}^{b}+S_{c\mu}{}^{\nu}\epsilon_{a}{}^{bc})W_{b\nu}=\rmi\Psibar\tau_{a}\gamma_{\mu}\spart\Psi-mJ_{a\mu}
\end{eqnarray}
Multiply by $\Psibar\gamma_{5}\gamma_{\mu}$:
\begin{eqnarray}
(g/2)\rmi\Sdual^{b}{}_{\mu}{}^{\nu}W_{b\nu}=(\rmi/2)(\Psibar\lspart\gamma_{5}\gamma_{\mu}\Psi-\Psibar\gamma_{5}\gamma_{\mu}\spart\Psi)+mK_{0\mu}
\end{eqnarray}
\begin{eqnarray}
\Psibar\lspart\gamma_{5}\gamma_{\mu}\Psi=-g\rmi K^{b}W_{b\mu}-\Psibar\gamma_{5}\gamma_{\mu}\spart\Psi
\end{eqnarray}
\begin{eqnarray}
(g/2)(K^{b}\delta_{\mu}{}^{\nu}-\rmi\Sdual^{b}{}_{\mu}{}^{\nu})W_{b\nu}=\rmi\Psibar\gamma_{5}\gamma_{\mu}\spart\Psi-mK_{0\mu}
\end{eqnarray}
Multiply by $\Psibar\tau_{a}\gamma_{5}\gamma_{\mu}$:
\begin{eqnarray}
\fl\Psibar\lspart\tau_{a}\gamma_{5}\gamma_{\mu}\Psi=-\rmi g(K_{0}\delta_{a}{}^{b}\delta_{\mu}{}^{\nu}+\Sdual_{c\mu}{}^{\nu}\epsilon_{a}{}^{bc})W_{b\nu}-\Psibar\tau_{a}\gamma_{5}\gamma_{\mu}\spart\Psi
\end{eqnarray}
\begin{eqnarray}
\fl\rmi(g/2)(K_{c}\delta_{\mu}{}^{\nu}\epsilon_{a}{}^{bc}-\Sdual_{0\mu}{}^{\nu}\delta_{a}{}^{b})W_{b\nu}=(\rmi/2)(\Psibar\tau_{a}\gamma_{5}\gamma_{\mu}\spart\Psi-\Psibar\lspart\tau_{a}\gamma_{5}\gamma_{\mu}\Psi)-mK_{a\mu}
\end{eqnarray}
\begin{eqnarray}
\fl(g/2)(K_{0}\delta_{\mu}{}^{\nu}\delta_{a}{}^{b}+\rmi K_{c}\delta_{\mu}{}^{\nu}\epsilon_{a}{}^{bc}-\rmi\Sdual_{0\mu}{}^{\nu}\delta_{a}{}^{b}+\Sdual_{c\mu}{}^{\nu}\epsilon_{a}{}^{bc})W_{b\nu}=\rmi\Psibar\tau_{a}\gamma_{5}\gamma_{\mu}\spart\Psi-mK_{a\mu}
\end{eqnarray}
Multiply by $\Psibar\sigma_{\rho\epsilon}$:
\begin{eqnarray}
\Psibar\lspart\sigma_{\rho\epsilon}\Psi=\Psibar\sigma_{\rho\epsilon}\spart\Psi-(\delta_{\epsilon}{}^{\nu}\delta_{\rho}{}^{\sigma}-\delta_{\rho}{}^{\nu}\delta_{\epsilon}{}^{\sigma})gJ^{b}{}_{\sigma}W_{b\nu}
\end{eqnarray}
\begin{eqnarray}
(g/2)\epsilon_{\rho\epsilon}{}^{\nu\sigma}K^{b}{}_{\sigma}W_{b\nu}=(\rmi/2)(\Psibar\sigma_{\rho\epsilon}\spart\Psi+\Psibar\lspart\sigma_{\rho\epsilon}\Psi)-mS_{0\rho\epsilon}
\end{eqnarray}
\begin{eqnarray}
(g/2)[\epsilon_{\rho\epsilon}{}^{\nu\sigma}K^{b}{}_{\sigma}+\rmi(\delta_{\epsilon}{}^{\nu}\delta_{\rho}{}^{\sigma}-\delta_{\rho}{}^{\nu}\delta_{\epsilon}{}^{\sigma})J^{b}{}_{\sigma}]W_{b\nu}=\rmi\Psibar\sigma_{\rho\epsilon}\spart\Psi-mS_{0\rho\epsilon}
\end{eqnarray}
Multiply by $\Psibar\tau_{a}\sigma_{\rho\epsilon}$:
\begin{eqnarray}
\fl(g/2)[\epsilon_{\rho\epsilon}{}^{\nu\sigma}K_{0\sigma}\delta_{a}{}^{b}+\rmi(\delta_{\rho}{}^{\sigma}\delta_{\epsilon}{}^{\nu}-\delta_{\epsilon}{}^{\sigma}\delta_{\rho}{}^{\nu})\rmi\epsilon_{a}{}^{bc}J_{c\sigma}]W_{b\nu} \nonumber \\
=(\rmi/2)(\Psibar\sigma_{\rho\epsilon}\tau_{a}\spart\Psi-\Psibar\lspart\sigma_{\rho\epsilon}\tau_{a}\Psi)-mS_{a\rho\epsilon}
\end{eqnarray}
\begin{eqnarray}
\fl\Psibar\lspart\sigma_{\rho\epsilon}\tau_{a}\Psi=g[(\delta_{\rho}{}^{\sigma}\delta_{\epsilon}{}^{\nu}-\delta_{\epsilon}{}^{\sigma}\delta_{\rho}{}^{\nu})J_{0\sigma}\delta_{a}{}^{b}+\epsilon_{\rho\epsilon}{}^{\nu\sigma}\epsilon_{a}{}^{bc}K_{c\sigma}]W_{b\nu}-\Psibar\sigma_{\rho\epsilon}\tau_{a}\spart\Psi
\end{eqnarray}
\begin{eqnarray}
\fl(g/2)[\epsilon_{\rho\epsilon}{}^{\nu\sigma}(K_{0\sigma}\delta_{a}{}^{b}+\rmi K_{c\sigma}\epsilon_{a}{}^{bc})+\rmi(\delta_{\rho}{}^{\sigma}\delta_{\epsilon}{}^{\nu}-\delta_{\epsilon}{}^{\sigma}\delta_{\rho}{}^{\nu})(J_{0\sigma}\delta_{a}{}^{b}+\rmi J_{c\sigma}\epsilon_{a}{}^{bc})]W_{b\nu} \nonumber \\
=\rmi\Psibar\sigma_{\rho\epsilon}\tau_{a}\spart\Psi-mS_{a\rho\epsilon}
\end{eqnarray}
Multiply by $\Psibar\gamma_{5}\sigma_{\rho\epsilon}$:
\begin{eqnarray}
\fl\Psibar\lspart\gamma_{5}\sigma_{\rho\epsilon}\Psi=-\Psibar\gamma_{5}\sigma_{\rho\epsilon}\spart\Psi-\rmi\epsilon_{\rho\epsilon}{}^{\nu\sigma}gJ^{b}{}_{\sigma}W_{b\nu}
\end{eqnarray}
\begin{eqnarray}
\fl(g/2)\rmi(\delta_{\epsilon}{}^{\nu}\delta_{\rho}{}^{\sigma}-\delta_{\rho}{}^{\nu}\delta_{\epsilon}{}^{\sigma})K^{b}{}_{\sigma}W_{b\nu}=(\rmi/2)(\Psibar\gamma_{5}\sigma_{\rho\epsilon}\spart\Psi-\Psibar\lspart\gamma_{5}\sigma_{\rho\epsilon}\Psi)-m\Sdual_{0\rho\epsilon}
\end{eqnarray}
\begin{eqnarray}
\fl(g/2)[\epsilon_{\rho\epsilon}{}^{\nu\sigma}J^{b}{}_{\sigma}+\rmi(\delta_{\epsilon}{}^{\nu}\delta_{\rho}{}^{\sigma}-\delta_{\rho}{}^{\nu}\delta_{\epsilon}{}^{\sigma})K^{b}{}_{\sigma}]W_{b\nu}=\rmi\Psibar\gamma_{5}\sigma_{\rho\epsilon}\spart\Psi-m\Sdual_{0\rho\epsilon}
\end{eqnarray}
Multiply by $\Psibar\tau_{a}\gamma_{5}\sigma_{\rho\epsilon}$:
\begin{eqnarray}
\fl(g/2)[\rmi(\delta_{\rho}{}^{\sigma}\delta_{\epsilon}{}^{\nu}-\delta_{\epsilon}{}^{\sigma}\delta_{\rho}{}^{\nu})K_{0\sigma}\delta_{a}{}^{b}+\rmi\epsilon_{\rho\epsilon}{}^{\nu\sigma}\epsilon_{a}{}^{bc}J_{c\sigma}]W_{b\nu} \nonumber \\
=(\rmi/2)(\Psibar\gamma_{5}\sigma_{\rho\epsilon}\tau_{a}\spart\Psi-\Psibar\lspart\gamma_{5}\sigma_{\rho\epsilon}\tau_{a}\Psi)-m\Sdual_{a\rho\epsilon}
\end{eqnarray}
\begin{eqnarray}
\fl\Psibar\lspart\gamma_{5}\sigma_{\rho\epsilon}\tau_{a}\Psi=g[(\delta_{\rho}{}^{\sigma}\delta_{\epsilon}{}^{\nu}-\delta_{\epsilon}{}^{\sigma}\delta_{\rho}{}^{\nu})\rmi\epsilon_{a}{}^{bc}K_{c\sigma}-\rmi\epsilon_{\rho\epsilon}{}^{\nu\sigma}J_{0\sigma}\delta_{a}{}^{b}]W_{b\nu}-\Psibar\gamma_{5}\sigma_{\rho\epsilon}\tau_{a}\spart\Psi
\end{eqnarray}
\begin{eqnarray}
\fl(g/2)[\epsilon_{\rho\epsilon}{}^{\nu\sigma}(J_{0\sigma}\delta_{a}{}^{b}+\rmi J_{c\sigma}\epsilon_{a}{}^{bc})+\rmi(\delta_{\rho}{}^{\sigma}\delta_{\epsilon}{}^{\nu}-\delta_{\epsilon}{}^{\sigma}\delta_{\rho}{}^{\nu})(K_{0\sigma}\delta_{a}{}^{b}+\rmi K_{c\sigma}\epsilon_{a}{}^{bc})]W_{b\nu} \nonumber \\
=\rmi\Psibar\gamma_{5}\sigma_{\rho\epsilon}\tau_{a}\spart\Psi-m\Sdual_{a\rho\epsilon}
\end{eqnarray}

\section{Fierz Identities for J-K Lorentz Vector Current Products}
Here follow the general Fierz identities for Lorentz vector currents $J_{i}{}^{\mu}$, multiplied by the dual $K_{j}{}^{\nu}$. There are four different types of Pauli term arrangements: $a-b$, $0-a$, $a-0$ and $0-0$ which indicate the $i=0,1,2,3$ index in $\tau_{i}$ for the left and right terms in the product $J_{i}{}^{\mu}K_{j}{}^{\nu}$. We treat the $i=0$ and $i=a=b=1,2,3$ Pauli indices separately due to their slightly different algebraic properties. For each case, we write out the full Fierz expansion in spinor form, then expand Pauli and Dirac matrix products in terms of irreducible elements of the Pauli and Dirac algebra using identities from appendix A. The final Fierz identities are written in the non-abelian current notation defined in Section 3.

\vspace{5mm}

\noindent $a-b$ case:
\begin{eqnarray}
\fl J_{a}{}^{\mu}K_{b}{}^{\nu}=\Psibar\tau_{a}\gamma^{\mu}(\Psi\Psibar)\tau_{b}\gamma_{5}\gamma^{\nu}\Psi \nonumber \\
=-(1/8)J_{i}\Psibar\gamma_{5}\gamma^{\mu}\gamma^{\nu}\tau_{a}\tau^{i}\tau_{b}\Psi+(1/8)J_{i\sigma}\Psibar\gamma_{5}\gamma^{\mu}\gamma^{\sigma}\gamma^{\nu}\tau_{a}\tau^{i}\tau_{b}\Psi \nonumber \\
\qquad-(1/16)S_{i\sigma\epsilon}\Psibar\gamma_{5}\gamma^{\mu}\sigma^{\sigma\epsilon}\gamma^{\nu}\tau_{a}\tau^{i}\tau_{b}\Psi \nonumber \\
\qquad+(1/8)K_{i\sigma}\Psibar\gamma^{\mu}\gamma^{\sigma}\gamma^{\nu}\tau_{a}\tau^{i}\tau_{b}\Psi+(1/8)K_{i}\Psibar\gamma^{\mu}\gamma^{\nu}\tau_{a}\tau^{i}\tau_{b}\Psi \nonumber \\
=-(1/8)J_{0}\Psibar\gamma_{5}[\eta^{\mu\nu}-\rmi\sigma^{\mu\nu}][\delta_{ab}+\rmi\epsilon_{ab}{}^{d}\tau_{d}]\Psi \nonumber \\
\qquad-(1/8)J_{c}\Psibar\gamma_{5}[\eta^{\mu\nu}-\rmi\sigma^{\mu\nu}][\tau_{a}\delta_{b}{}^{c}+\tau_{b}\delta_{a}{}^{c}-\tau^{c}\delta_{ab}-\rmi\epsilon_{ab}{}^{c}]\Psi \nonumber \\
\qquad+(1/8)J_{0\sigma}\Psibar\gamma_{5}[\eta^{\mu\sigma}\gamma^{\nu}+\eta^{\sigma\nu}\gamma^{\mu}-\eta^{\mu\nu}\gamma^{\sigma}-\rmi\epsilon^{\mu\sigma\nu\lambda}\gamma_{5}\gamma_{\lambda}][\delta_{ab}+\rmi\epsilon_{ab}{}^{d}\tau_{d}]\Psi \nonumber \\
\qquad+(1/8)J_{c\sigma}\Psibar\gamma_{5}[\eta^{\mu\sigma}\gamma^{\nu}+\eta^{\sigma\nu}\gamma^{\mu}-\eta^{\mu\nu}\gamma^{\sigma}-\rmi\epsilon^{\mu\sigma\nu\lambda}\gamma_{5}\gamma_{\lambda}][\tau_{a}\delta_{b}{}^{c}+\tau_{b}\delta_{a}{}^{c} \nonumber \\
\qquad\qquad-\tau^{c}\delta_{ab}-\rmi\epsilon_{ab}{}^{c}]\Psi \nonumber \\
\qquad-(1/16)S_{0\sigma\epsilon}\Psibar\gamma_{5}[\rmi\eta^{\epsilon\nu}\eta^{\mu\sigma}-\rmi\eta^{\sigma\nu}\eta^{\mu\epsilon}+\eta^{\epsilon\nu}\sigma^{\mu\sigma}-\eta^{\sigma\nu}\sigma^{\mu\epsilon}-\epsilon^{\sigma\epsilon\nu\mu}\gamma_{5} \nonumber \\
\qquad\qquad+\rmi\epsilon^{\sigma\epsilon\nu\lambda}\gamma_{5}\sigma^{\mu}{}_{\lambda}][\delta_{ab}+\rmi\epsilon_{ab}{}^{d}\tau_{d}]\Psi \nonumber \\
\qquad-(1/16)S_{c\sigma\epsilon}\Psibar\gamma_{5}[\rmi\eta^{\epsilon\nu}\eta^{\mu\sigma}-\rmi\eta^{\sigma\nu}\eta^{\mu\epsilon}+\eta^{\epsilon\nu}\sigma^{\mu\sigma}-\eta^{\sigma\nu}\sigma^{\mu\epsilon}-\epsilon^{\sigma\epsilon\nu\mu}\gamma_{5} \nonumber \\
\qquad\qquad+\rmi\epsilon^{\sigma\epsilon\nu\lambda}\gamma_{5}\sigma^{\mu}{}_{\lambda}][\tau_{a}\delta_{b}{}^{c}+\tau_{b}\delta_{a}{}^{c}-\tau^{c}\delta_{ab}-\rmi\epsilon_{ab}{}^{c}]\Psi \nonumber \\
\qquad+(1/8)K_{0\sigma}\Psibar[\eta^{\mu\sigma}\gamma^{\nu}+\eta^{\sigma\nu}\gamma^{\mu}-\eta^{\mu\nu}\gamma^{\sigma}-\rmi\epsilon^{\mu\sigma\nu\lambda}\gamma_{5}\gamma_{\lambda}][\delta_{ab}+\rmi\epsilon_{ab}{}^{d}\tau_{d}]\Psi \nonumber \\
\qquad+(1/8)K_{c\sigma}\Psibar[\eta^{\mu\sigma}\gamma^{\nu}+\eta^{\sigma\nu}\gamma^{\mu}-\eta^{\mu\nu}\gamma^{\sigma}-\rmi\epsilon^{\mu\sigma\nu\lambda}\gamma_{5}\gamma_{\lambda}][\tau_{a}\delta_{b}{}^{c}+\tau_{b}\delta_{a}{}^{c} \nonumber \\
\qquad\qquad-\tau^{c}\delta_{ab}-\rmi\epsilon_{ab}{}^{c}]\Psi \nonumber \\
\qquad+(1/8)K_{0}\Psibar[\eta^{\mu\nu}-\rmi\sigma^{\mu\nu}][\delta_{ab}+\rmi\epsilon_{ab}{}^{d}\tau_{d}]\Psi \nonumber \\
\qquad+(1/8)K_{c}\Psibar[\eta^{\mu\nu}-\rmi\sigma^{\mu\nu}][\tau_{a}\delta_{b}{}^{c}+\tau_{b}\delta_{a}{}^{c}-\tau^{c}\delta_{ab}-\rmi\epsilon_{ab}{}^{c}]\Psi \nonumber
\end{eqnarray}
\begin{eqnarray}\label{JK Fierz a-b case}
\fl J_{a}{}^{\mu}K_{b}{}^{\nu}=(1/4)[\rmi J_{a}\Sdual_{b}{}^{\mu\nu}+\rmi J_{b}\Sdual_{a}{}^{\mu\nu}-\rmi K_{a}S_{b}{}^{\mu\nu}-\rmi K_{b}S_{a}{}^{\mu\nu}+J_{a}{}^{\mu}K_{b}{}^{\nu} \nonumber \\
+J_{a}{}^{\nu}K_{b}{}^{\mu}+J_{b}{}^{\mu}K_{a}{}^{\nu}+J_{b}{}^{\nu}K_{a}{}^{\mu}-J_{a\sigma}K_{b}{}^{\sigma}\eta^{\mu\nu}-J_{b\sigma}K_{a}{}^{\sigma}\eta^{\mu\nu} \nonumber \\
+\delta_{ab}(\rmi J_{0}\Sdual_{0}{}^{\mu\nu}-\rmi J_{c}\Sdual^{c\mu\nu}-\rmi K_{0}S_{0}{}^{\mu\nu}+\rmi K_{c}S^{c\mu\nu}+J_{0}{}^{\mu}K_{0}{}^{\nu} \nonumber \\
+J_{0}{}^{\nu}K_{0}{}^{\mu}-J_{c}{}^{\mu}K^{c\nu}-J_{c}{}^{\nu}K^{c\mu}-J_{0\sigma}K_{0}{}^{\sigma}\eta^{\mu\nu}+J_{c\sigma}K^{c\sigma}\eta^{\mu\nu})] \nonumber \\
+(1/4)\epsilon_{ab}{}^{c}[-\rmi J_{0}K_{c}\eta^{\mu\nu}+\rmi K_{0}J_{c}\eta^{\mu\nu}+J_{c\sigma}J_{0\lambda}\epsilon^{\mu\nu\sigma\lambda}+K_{c\sigma}K_{0\lambda}\epsilon^{\mu\nu\sigma\lambda} \nonumber \\
+(1/2)\rmi(-S_{0}{}^{\mu}{}_{\sigma}\Sdual_{c}{}^{\sigma\nu}-S_{0}{}^{\nu}{}_{\sigma}\Sdual_{c}{}^{\sigma\mu}+S_{c}{}^{\mu}{}_{\sigma}\Sdual_{0}{}^{\sigma\nu}+S_{c}{}^{\nu}{}_{\sigma}\Sdual_{0}{}^{\sigma\mu})].
\end{eqnarray}
$a-0$ case:
\begin{eqnarray}
\fl J_{a}{}^{\mu}K_{0}{}^{\nu}=\Psibar\tau_{a}\gamma^{\mu}(\Psi\Psibar)\gamma_{5}\gamma^{\nu}\Psi \nonumber \\
=-(1/8)J_{i}\Psibar\gamma_{5}\gamma^{\mu}\gamma^{\nu}\tau_{a}\tau^{i}\Psi+(1/8)J_{i\sigma}\Psibar\gamma_{5}\gamma^{\mu}\gamma^{\sigma}\gamma^{\nu}\tau_{a}\tau^{i}\Psi \nonumber \\
\qquad-(1/16)S_{i\sigma\epsilon}\Psibar\gamma_{5}\gamma^{\mu}\sigma^{\sigma\epsilon}\gamma^{\nu}\tau_{a}\tau^{i}\Psi \nonumber \\
\qquad+(1/8)K_{i\sigma}\Psibar\gamma^{\mu}\gamma^{\sigma}\gamma^{\nu}\tau_{a}\tau^{i}\Psi+(1/8)K_{i}\Psibar\gamma^{\mu}\gamma^{\nu}\tau_{a}\tau^{i}\Psi \nonumber \\
=-(1/8)J_{0}\Psibar\gamma_{5}[\eta^{\mu\nu}-\rmi\sigma^{\mu\nu}]\tau_{a}\Psi-(1/8)J_{c}\Psibar\gamma_{5}[\eta^{\mu\nu}-\rmi\sigma^{\mu\nu}][\delta_{a}{}^{c}+\rmi\epsilon_{a}{}^{cd}\tau_{d}]\Psi \nonumber \\
\qquad+(1/8)J_{0\sigma}\Psibar\gamma_{5}[\eta^{\mu\sigma}\gamma^{\nu}+\eta^{\sigma\nu}\gamma^{\mu}-\eta^{\mu\nu}\gamma^{\sigma}-\rmi\epsilon^{\mu\sigma\nu\lambda}\gamma_{5}\gamma_{\lambda}]\tau_{a}\Psi \nonumber \\
\qquad+(1/8)J_{c\sigma}\Psibar\gamma_{5}[\eta^{\mu\sigma}\gamma^{\nu}+\eta^{\sigma\nu}\gamma^{\mu}-\eta^{\mu\nu}\gamma^{\sigma}-\rmi\epsilon^{\mu\sigma\nu\lambda}\gamma_{5}\gamma_{\lambda}][\delta_{a}{}^{c}+\rmi\epsilon_{a}{}^{cd}\tau_{d}]\Psi \nonumber \\
\qquad-(1/16)S_{0\sigma\epsilon}\Psibar\gamma_{5}[\rmi\eta^{\epsilon\nu}\eta^{\mu\sigma}-\rmi\eta^{\sigma\nu}\eta^{\mu\epsilon}+\eta^{\epsilon\nu}\sigma^{\mu\sigma}-\eta^{\sigma\nu}\sigma^{\mu\epsilon}-\epsilon^{\sigma\epsilon\nu\mu}\gamma_{5} \nonumber \\
\qquad\qquad+\rmi\epsilon^{\sigma\epsilon\nu\lambda}\gamma_{5}\sigma^{\mu}{}_{\lambda}]\tau_{a}\Psi \nonumber \\
\qquad-(1/16)S_{c\sigma\epsilon}\Psibar\gamma_{5}[\rmi\eta^{\epsilon\nu}\eta^{\mu\sigma}-\rmi\eta^{\sigma\nu}\eta^{\mu\epsilon}+\eta^{\epsilon\nu}\sigma^{\mu\sigma}-\eta^{\sigma\nu}\sigma^{\mu\epsilon}-\epsilon^{\sigma\epsilon\nu\mu}\gamma_{5} \nonumber \\
\qquad\qquad+\rmi\epsilon^{\sigma\epsilon\nu\lambda}\gamma_{5}\sigma^{\mu}{}_{\lambda}][\delta_{a}{}^{c}+\rmi\epsilon_{a}{}^{cd}\tau_{d}]\Psi \nonumber \\
\qquad+(1/8)K_{0\sigma}\Psibar[\eta^{\mu\sigma}\gamma^{\nu}+\eta^{\sigma\nu}\gamma^{\mu}-\eta^{\mu\nu}\gamma^{\sigma}-\rmi\epsilon^{\mu\sigma\nu\lambda}\gamma_{5}\gamma_{\lambda}]\tau_{a}\Psi \nonumber \\
\qquad+(1/8)K_{c\sigma}\Psibar[\eta^{\mu\sigma}\gamma^{\nu}+\eta^{\sigma\nu}\gamma^{\mu}-\eta^{\mu\nu}\gamma^{\sigma}-\rmi\epsilon^{\mu\sigma\nu\lambda}\gamma_{5}\gamma_{\lambda}][\delta_{a}{}^{c}+\rmi\epsilon_{a}{}^{cd}\tau_{d}]\Psi \nonumber \\
\qquad+(1/8)K_{0}\Psibar[\eta^{\mu\nu}-\rmi\sigma^{\mu\nu}]\tau_{a}\Psi \nonumber \\
\qquad+(1/8)K_{c}\Psibar[\eta^{\mu\nu}-\rmi\sigma^{\mu\nu}][\delta_{a}{}^{c}+\rmi\epsilon_{a}{}^{cd}\tau_{d}]\Psi \nonumber
\end{eqnarray}
\begin{eqnarray}\label{JK Fierz a-0 case}
\fl J_{a}{}^{\mu}K_{0}{}^{\nu}=(1/4)[\rmi J_{0}\Sdual_{a}{}^{\mu\nu}+\rmi J_{a}\Sdual_{0}{}^{\mu\nu}-\rmi K_{0}S_{a}{}^{\mu\nu}-\rmi K_{a}S_{0}{}^{\mu\nu}+J_{0}{}^{\mu}K_{a}{}^{\nu} \nonumber \\
+J_{0}{}^{\nu}K_{a}{}^{\mu}+J_{a}{}^{\mu}K_{0}{}^{\nu}+J_{a}{}^{\nu}K_{0}{}^{\mu}-J_{0\sigma}K_{a}{}^{\sigma}\eta^{\mu\nu}-J_{a\sigma}K_{0}{}^{\sigma}\eta^{\mu\nu}] \nonumber \\
-(1/4)\epsilon_{a}{}^{cd}[\rmi J_{c}K_{d}\eta^{\mu\nu}+(1/2)J_{c\sigma}J_{d\lambda}\epsilon^{\mu\nu\sigma\lambda}+(1/2)K_{c\sigma}K_{d\lambda}\epsilon^{\mu\nu\sigma\lambda} \nonumber \\
-(1/2)\rmi(S_{d}{}^{\mu}{}_{\sigma}\Sdual_{c}{}^{\sigma\nu}+S_{d}{}^{\nu}{}_{\sigma}\Sdual_{c}{}^{\sigma\mu})]
\end{eqnarray}
$0-a$ case:
\begin{eqnarray}
\fl J_{0}{}^{\mu}K_{a}{}^{\nu}=\Psibar\gamma^{\mu}(\Psi\Psibar)\tau_{a}\gamma_{5}\gamma^{\nu}\Psi \nonumber \\
=-(1/8)J_{i}\Psibar\gamma_{5}\gamma^{\mu}\gamma^{\nu}\tau^{i}\tau_{a}\Psi+(1/8)J_{i\sigma}\Psibar\gamma_{5}\gamma^{\mu}\gamma^{\sigma}\gamma^{\nu}\tau^{i}\tau_{a}\Psi \nonumber \\
\qquad-(1/16)S_{i\sigma\epsilon}\Psibar\gamma_{5}\gamma^{\mu}\sigma^{\sigma\epsilon}\gamma^{\nu}\tau^{i}\tau_{a}\Psi \nonumber \\
\qquad+(1/8)K_{i\sigma}\Psibar\gamma^{\mu}\gamma^{\sigma}\gamma^{\nu}\tau^{i}\tau_{a}\Psi+(1/8)K_{i}\Psibar\gamma^{\mu}\gamma^{\nu}\tau^{i}\tau_{a}\Psi\nonumber \\
=-(1/8)J_{0}\Psibar\gamma_{5}[\eta^{\mu\nu}-\rmi\sigma^{\mu\nu}]\tau_{a}\Psi-(1/8)J_{c}\Psibar\gamma_{5}[\eta^{\mu\nu}-\rmi\sigma^{\mu\nu}][\delta_{a}{}^{c}-\rmi\epsilon_{a}{}^{cd}\tau_{d}]\Psi \nonumber \\
\qquad+(1/8)J_{0\sigma}\Psibar\gamma_{5}[\eta^{\mu\sigma}\gamma^{\nu}+\eta^{\sigma\nu}\gamma^{\mu}-\eta^{\mu\nu}\gamma^{\sigma}-\rmi\epsilon^{\mu\sigma\nu\lambda}\gamma_{5}\gamma_{\lambda}]\tau_{a}\Psi \nonumber \\
\qquad+(1/8)J_{c\sigma}\Psibar\gamma_{5}[\eta^{\mu\sigma}\gamma^{\nu}+\eta^{\sigma\nu}\gamma^{\mu}-\eta^{\mu\nu}\gamma^{\sigma}-\rmi\epsilon^{\mu\sigma\nu\lambda}\gamma_{5}\gamma_{\lambda}][\delta_{a}{}^{c}-\rmi\epsilon_{a}{}^{cd}\tau_{d}]\Psi \nonumber \\
\qquad-(1/16)S_{0\sigma\epsilon}\Psibar\gamma_{5}[\rmi\eta^{\epsilon\nu}\eta^{\mu\sigma}-\rmi\eta^{\sigma\nu}\eta^{\mu\epsilon}+\eta^{\epsilon\nu}\sigma^{\mu\sigma}-\eta^{\sigma\nu}\sigma^{\mu\epsilon}-\epsilon^{\sigma\epsilon\nu\mu}\gamma_{5} \nonumber \\
\qquad\qquad+\rmi\epsilon^{\sigma\epsilon\nu\lambda}\gamma_{5}\sigma^{\mu}{}_{\lambda}]\tau_{a}\Psi \nonumber \\
\qquad-(1/16)S_{c\sigma\epsilon}\Psibar\gamma_{5}[\rmi\eta^{\epsilon\nu}\eta^{\mu\sigma}-\rmi\eta^{\sigma\nu}\eta^{\mu\epsilon}+\eta^{\epsilon\nu}\sigma^{\mu\sigma}-\eta^{\sigma\nu}\sigma^{\mu\epsilon}-\epsilon^{\sigma\epsilon\nu\mu}\gamma_{5} \nonumber \\
\qquad\qquad+\rmi\epsilon^{\sigma\epsilon\nu\lambda}\gamma_{5}\sigma^{\mu}{}_{\lambda}][\delta_{a}{}^{c}-\rmi\epsilon_{a}{}^{cd}\tau_{d}]\Psi \nonumber \\
\qquad+(1/8)K_{0\sigma}\Psibar[\eta^{\mu\sigma}\gamma^{\nu}+\eta^{\sigma\nu}\gamma^{\mu}-\eta^{\mu\nu}\gamma^{\sigma}-\rmi\epsilon^{\mu\sigma\nu\lambda}\gamma_{5}\gamma_{\lambda}]\tau_{a}\Psi \nonumber \\
\qquad+(1/8)K_{c\sigma}\Psibar[\eta^{\mu\sigma}\gamma^{\nu}+\eta^{\sigma\nu}\gamma^{\mu}-\eta^{\mu\nu}\gamma^{\sigma}-\rmi\epsilon^{\mu\sigma\nu\lambda}\gamma_{5}\gamma_{\lambda}][\delta_{a}{}^{c}-\rmi\epsilon_{a}{}^{cd}\tau_{d}]\Psi \nonumber \\
\qquad+(1/8)K_{0}\Psibar[\eta^{\mu\nu}-\rmi\sigma^{\mu\nu}]\tau_{a}\Psi \nonumber \\
\qquad+(1/8)K_{c}\Psibar[\eta^{\mu\nu}-\rmi\sigma^{\mu\nu}][\delta_{a}{}^{c}-\rmi\epsilon_{a}{}^{cd}\tau_{d}]\Psi \nonumber
\end{eqnarray}
\begin{eqnarray}\label{JK Fierz 0-a case}
\fl J_{0}{}^{\mu}K_{a}{}^{\nu}=(1/4)[\rmi J_{0}\Sdual_{a}{}^{\mu\nu}+\rmi J_{a}\Sdual_{0}{}^{\mu\nu}-\rmi K_{0}S_{a}{}^{\mu\nu}-\rmi K_{a}S_{0}{}^{\mu\nu}+J_{0}{}^{\mu}K_{a}{}^{\nu} \nonumber \\
+J_{0}{}^{\nu}K_{a}{}^{\mu}+J_{a}{}^{\mu}K_{0}{}^{\nu}+J_{a}{}^{\nu}K_{0}{}^{\mu}-J_{0\sigma}K_{a}{}^{\sigma}\eta^{\mu\nu}-J_{a\sigma}K_{0}{}^{\sigma}\eta^{\mu\nu}] \nonumber \\
+(1/4)\epsilon_{a}{}^{cd}[\rmi J_{c}K_{d}\eta^{\mu\nu}+(1/2)J_{c\sigma}J_{d\lambda}\epsilon^{\mu\nu\sigma\lambda}+(1/2)K_{c\sigma}K_{d\lambda}\epsilon^{\mu\nu\sigma\lambda} \nonumber \\
-(1/2)\rmi(S_{d}{}^{\mu}{}_{\sigma}\Sdual_{c}{}^{\sigma\nu}+S_{d}{}^{\nu}{}_{\sigma}\Sdual_{c}{}^{\sigma\mu})]
\end{eqnarray}
$0-0$ case:
\begin{eqnarray}
\fl J_{0}{}^{\mu}K_{0}{}^{\nu}=\Psibar\gamma^{\mu}(\Psi\Psibar)\gamma_{5}\gamma^{\nu}\Psi \nonumber \\
=-(1/8)J_{i}\Psibar\gamma_{5}\gamma^{\mu}\gamma^{\nu}\tau^{i}\Psi+(1/8)J_{i\sigma}\Psibar\gamma_{5}\gamma^{\mu}\gamma^{\sigma}\gamma^{\nu}\tau^{i}\Psi \nonumber \\
\qquad-(1/16)S_{i\sigma\epsilon}\Psibar\gamma_{5}\gamma^{\mu}\sigma^{\sigma\epsilon}\gamma^{\nu}\tau^{i}\Psi \nonumber \\
\qquad+(1/8)K_{i\sigma}\Psibar\gamma^{\mu}\gamma^{\sigma}\gamma^{\nu}\tau^{i}\Psi+(1/8)K_{i}\Psibar\gamma^{\mu}\gamma^{\nu}\tau^{i}\Psi \nonumber \\
=-(1/8)J_{0}\Psibar\gamma_{5}[\eta^{\mu\nu}-\rmi\sigma^{\mu\nu}]\Psi-(1/8)J_{c}\Psibar\gamma_{5}[\eta^{\mu\nu}-\rmi\sigma^{\mu\nu}]\tau^{c}\Psi \nonumber \\
\qquad+(1/8)J_{0\sigma}\Psibar\gamma_{5}[\eta^{\mu\sigma}\gamma^{\nu}+\eta^{\sigma\nu}\gamma^{\mu}-\eta^{\mu\nu}\gamma^{\sigma}-\rmi\epsilon^{\mu\sigma\nu\lambda}\gamma_{5}\gamma_{\lambda}]\Psi \nonumber \\
\qquad+(1/8)J_{c\sigma}\Psibar\gamma_{5}[\eta^{\mu\sigma}\gamma^{\nu}+\eta^{\sigma\nu}\gamma^{\mu}-\eta^{\mu\nu}\gamma^{\sigma}-\rmi\epsilon^{\mu\sigma\nu\lambda}\gamma_{5}\gamma_{\lambda}]\tau^{c}\Psi \nonumber \\
\qquad-(1/16)S_{0\sigma\epsilon}\Psibar\gamma_{5}[\rmi\eta^{\epsilon\nu}\eta^{\mu\sigma}-\rmi\eta^{\sigma\nu}\eta^{\mu\epsilon}+\eta^{\epsilon\nu}\sigma^{\mu\sigma}-\eta^{\sigma\nu}\sigma^{\mu\epsilon}-\epsilon^{\sigma\epsilon\nu\mu}\gamma_{5} \nonumber \\
\qquad\qquad+\rmi\epsilon^{\sigma\epsilon\nu\lambda}\gamma_{5}\sigma^{\mu}{}_{\lambda}]\Psi \nonumber \\
\qquad-(1/16)S_{c\sigma\epsilon}\Psibar\gamma_{5}[\rmi\eta^{\epsilon\nu}\eta^{\mu\sigma}-\rmi\eta^{\sigma\nu}\eta^{\mu\epsilon}+\eta^{\epsilon\nu}\sigma^{\mu\sigma}-\eta^{\sigma\nu}\sigma^{\mu\epsilon}-\epsilon^{\sigma\epsilon\nu\mu}\gamma_{5} \nonumber \\
\qquad\qquad+\rmi\epsilon^{\sigma\epsilon\nu\lambda}\gamma_{5}\sigma^{\mu}{}_{\lambda}]\tau^{c}\Psi \nonumber \\
\qquad+(1/8)K_{0\sigma}\Psibar[\eta^{\mu\sigma}\gamma^{\nu}+\eta^{\sigma\nu}\gamma^{\mu}-\eta^{\mu\nu}\gamma^{\sigma}-\rmi\epsilon^{\mu\sigma\nu\lambda}\gamma_{5}\gamma_{\lambda}]\Psi \nonumber \\
\qquad+(1/8)K_{c\sigma}\Psibar[\eta^{\mu\sigma}\gamma^{\nu}+\eta^{\sigma\nu}\gamma^{\mu}-\eta^{\mu\nu}\gamma^{\sigma}-\rmi\epsilon^{\mu\sigma\nu\lambda}\gamma_{5}\gamma_{\lambda}]\tau^{c}\Psi \nonumber \\
\qquad+(1/8)K_{0}\Psibar[\eta^{\mu\nu}-\rmi\sigma^{\mu\nu}]\Psi+(1/8)K_{c}\Psibar[\eta^{\mu\nu}-\rmi\sigma^{\mu\nu}]\tau^{c}\Psi \nonumber
\end{eqnarray}
\begin{eqnarray}\label{JK Fierz 0-0 case}
\fl J_{0}{}^{\mu}K_{0}{}^{\nu}=(1/4)[\rmi J_{0}\Sdual_{0}{}^{\mu\nu}+\rmi J_{c}\Sdual^{c\mu\nu}-\rmi K_{0}S_{0}{}^{\mu\nu}-\rmi K_{c}S^{c\mu\nu}+J_{0}{}^{\mu}K_{0}{}^{\nu}+J_{0}{}^{\nu}K_{0}{}^{\mu} \nonumber \\
+J_{c}{}^{\mu}K^{c\nu}+J_{c}{}^{\nu}K^{c\mu}-J_{0\sigma}K_{0}{}^{\sigma}\eta^{\mu\nu}-J_{c\sigma}K^{c\sigma}\eta^{\mu\nu}]
\end{eqnarray}

\section{Fierz Identities for Rank-2 Skew Tensor Currents}
This appendix contains a more detailed version of the derivation of the Fierz identities for $S_{0}{}^{\mu\nu}$, $\Sdual_{0}{}^{\mu\nu}$, $S_{a}{}^{\mu\nu}$ and $\Sdual_{a}{}^{\mu\nu}$. Let us first derive expressions for $S_{0}{}^{\mu\nu}$ and $\Sdual_{0}{}^{\mu\nu}$, using the Fierz identities for $JK$ vector current products derived in appendix C. Consider the following expression, obtained from the Fierz identity (\ref{JK Fierz 0-0 case}) and subtracting the Pauli trace of (\ref{JK Fierz a-b case}):
\begin{eqnarray}
\fl J_{0}{}^{\mu}K_{0}{}^{\nu}-J_{a}{}^{\nu}K^{a\mu}=\rmi J_{0}\Sdual_{0}{}^{\mu\nu}-\rmi K_{0}S_{0}{}^{\mu\nu}-(1/2)J_{0}{}^{\mu}K_{0}{}^{\nu}-(1/2)J_{0}{}^{\nu}K_{0}{}^{\mu}+(1/2)J_{a}{}^{\mu}K^{a\nu} \nonumber \\
+(1/2)J_{a}{}^{\nu}K^{a\mu}+(1/2)J_{0\sigma}K_{0}{}^{\sigma}\eta^{\mu\nu}-(1/2)J_{a\sigma}K^{a\sigma}\eta^{\mu\nu}.
\end{eqnarray}
Note that we have used the Fierz forms that have no stand-alone Pauli vector triplet indices $a=1,2,3$. Similarly, using the same Fierz identities we can also form
\begin{eqnarray}
\fl J_{a}{}^{\mu}K^{a\nu}-J_{0}{}^{\nu}K_{0}{}^{\mu}=\rmi J_{0}\Sdual_{0}{}^{\mu\nu}-\rmi K_{0}S_{0}{}^{\mu\nu}+(1/2)J_{0}{}^{\mu}K_{0}{}^{\nu}+(1/2)J_{0}{}^{\nu}K_{0}{}^{\mu}-(1/2)J_{a}{}^{\mu}K^{a\nu} \nonumber \\
-(1/2)J_{a}{}^{\nu}K^{a\mu}-(1/2)J_{0\sigma}K_{0}{}^{\sigma}\eta^{\mu\nu}+(1/2)J_{a\sigma}K^{a\sigma}\eta^{\mu\nu}.
\end{eqnarray}
Adding these two equations, we get
\begin{eqnarray}\label{Lorentz antisymmetric subtraction term, trace over i}
J_{i}{}^{\mu}K^{i\nu}-J_{i}{}^{\nu}K^{i\mu}=2\rmi(J_{0}\Sdual_{0}{}^{\mu\nu}-K_{0}S_{0}{}^{\mu\nu})
\end{eqnarray}
which is antisymmetric in $\mu, \nu$. Another Fierz expression which is antisymmetric in $\mu, \nu$, is
\begin{eqnarray}\label{JK 0-0 case Levi-Civita contracted}
\fl\epsilon^{\mu\nu\rho\kappa}J_{0\rho}K_{0\kappa}=(1/4)\epsilon^{\mu\nu\rho\kappa}[\rmi J_{0}\Sdual_{0\rho\kappa}+\rmi J_{a}\Sdual^{a}{}_{\rho\kappa}-\rmi K_{0}S_{0\rho\kappa}-\rmi K_{a}S^{a}{}_{\rho\kappa}+J_{0\rho}K_{0\kappa}+J_{0\kappa}K_{0\rho} \nonumber \\
\qquad+J_{a\rho}K^{a}{}_{\kappa}+J_{a\kappa}K^{a}{}_{\rho}-J_{0\sigma}K_{0}{}^{\sigma}\eta_{\rho\kappa}-J_{a\sigma}K^{a\sigma}\eta_{\rho\kappa}] \nonumber \\
=-(1/8)(\epsilon^{\mu\nu\rho\kappa}\epsilon_{\rho\kappa\delta\tau})J_{0}S_{0}{}^{\delta\tau}-(1/8)(\epsilon^{\mu\nu\rho\kappa}\epsilon_{\rho\kappa\delta\tau})J_{a}S^{a\delta\tau}-(1/2)K_{0}\Sdual_{0}{}^{\mu\nu} \nonumber \\
\qquad-(1/2)K_{a}\Sdual^{a\mu\nu} \nonumber \\
=(1/2)J_{0}S_{0}{}^{\mu\nu}+(1/2)J_{a}S^{a\mu\nu}-(1/2)K_{0}\Sdual_{0}{}^{\mu\nu}-(1/2)K_{a}\Sdual^{a\mu\nu}
\end{eqnarray}
where we have first canceled out all of the terms from (\ref{JK Fierz 0-0 case}) symmetric in $\mu, \nu$, then applied the Levi-Civita double contraction identity
\begin{eqnarray}\label{Levi-Civita double contraction identity}
\epsilon^{\mu\nu\rho\sigma}\epsilon_{\mu\nu\rho'\sigma'}=2(\delta^{\rho}{}_{\sigma'}\delta^{\sigma}{}_{\rho'}-\delta^{\rho}{}_{\rho'}\delta^{\sigma}{}_{\sigma'})
\end{eqnarray}
in addition to the definition (\ref{S-dual definition}) to convert the $S$ terms to $\Sdual$, and vice-versa. The other term we can contract with the Levi-Civita tensor, with no stand-alone Pauli vector triplet indices is the trace over the Pauli index $a$. So after (again) canceling all symmetric terms from (\ref{JK Fierz a-b case}) and applying the Levi-Civita contraction identity, we get
\begin{eqnarray}\label{JK a-b case Levi-Civita contracted}
\fl\epsilon^{\mu\nu\rho\kappa}J_{a\rho}K^{a}{}_{\kappa}=(3/2)J_{0}S_{0}{}^{\mu\nu}-(1/2)J_{a}S^{a\mu\nu}-(3/2)K_{0}\Sdual_{0}{}^{\mu\nu}+(1/2)K_{a}\Sdual^{a\mu\nu}
\end{eqnarray}
Taking the sum of (\ref{JK 0-0 case Levi-Civita contracted}) and (\ref{JK a-b case Levi-Civita contracted}), we obtain
\begin{eqnarray}\label{Lorentz antisymmetric Levi-Civita contracted term, trace over i}
\epsilon^{\mu\nu\rho\kappa}J_{i\rho}K^{i}{}_{\kappa}=2(J_{0}S_{0}{}^{\mu\nu}-K_{0}\Sdual_{0}{}^{\mu\nu}).
\end{eqnarray}
Now take the combination of (\ref{Lorentz antisymmetric subtraction term, trace over i}) and (\ref{Lorentz antisymmetric Levi-Civita contracted term, trace over i})
\begin{eqnarray}
\fl J_{0}\epsilon^{\mu\nu\rho\kappa}J_{i\rho}K^{i}{}_{\kappa}-\rmi K_{0}(J_{i}{}^{\mu}K^{i\nu}-J_{i}{}^{\nu}K^{i\mu}) \nonumber \\
=2(J_{0}^{2}S_{0}{}^{\mu\nu}-J_{0}K_{0}\Sdual_{0}{}^{\mu\nu}+J_{0}K_{0}\Sdual_{0}{}^{\mu\nu}-K_{0}^{2}S_{0}{}^{\mu\nu}).
\end{eqnarray}
Canceling the middle two terms on the right-hand side and rearranging, we end up with one of our identities
\begin{eqnarray}\label{S0 Fierz identity appendix}
S_{0}{}^{\mu\nu}=(1/2)(J_{0}^{2}-K_{0}^{2})^{-1}[J_{0}\epsilon^{\mu\nu\rho\kappa}J_{i\rho}K^{i}{}_{\kappa}-\rmi K_{0}(J_{i}{}^{\mu}K^{i\nu}-J_{i}{}^{\nu}K^{i\mu})].
\end{eqnarray}
By comparison, the abelian version of the identity is \cite{Crawford-1985}, \cite{Kaempffer-1981},
\begin{eqnarray}
s^{\mu\nu}=(\sigma^{2}-\omega^{2})^{-1}[\sigma\epsilon^{\mu\nu\rho\kappa}j_{\rho}k_{\kappa}-\rmi\omega(j^{\mu}k^{\nu}-j^{\nu}k^{\mu})],
\end{eqnarray}
where we have used an alternate definition for the pseudoscalar, $\omega\equiv\psibar\gamma_{5}\psi$, as opposed to the often used $\psibar\rmi\gamma_{5}\psi$. Now to calculate the dual, take an alternate combination of (\ref{Lorentz antisymmetric subtraction term, trace over i}) and (\ref{Lorentz antisymmetric Levi-Civita contracted term, trace over i})
\begin{eqnarray}
\fl K_{0}\epsilon^{\mu\nu\rho\kappa}J_{i\rho}K^{i}{}_{\kappa}-\rmi J_{0}(J_{i}{}^{\mu}K^{i\nu}-J_{i}{}^{\nu}K^{i\mu}) \nonumber \\
=2(J_{0}K_{0}S_{0}{}^{\mu\nu}-K_{0}^{2}\Sdual_{0}{}^{\mu\nu}+J_{0}^{2}\Sdual_{0}{}^{\mu\nu}-J_{0}K_{0}S_{0}{}^{\mu\nu})
\end{eqnarray}
which immediately leads to
\begin{eqnarray}\label{Sdual0 Fierz identity appendix}
\Sdual_{0}{}^{\mu\nu}=(1/2)(J_{0}^{2}-K_{0}^{2})^{-1}[K_{0}\epsilon^{\mu\nu\rho\kappa}J_{i\rho}K^{i}{}_{\kappa}-\rmi J_{0}(J_{i}{}^{\mu}K^{i\nu}-J_{i}{}^{\nu}K^{i\mu})].
\end{eqnarray}
Let us now derive an expression for $S_{a}{}^{\mu\nu}$. We start by forming an expression from the $JK$ Lorentz vector current product Fierz identities, with a single Pauli vector triplet index, (\ref{JK Fierz a-0 case}) and (\ref{JK Fierz 0-a case}),
\begin{eqnarray}\label{Sum of a-0 and 0-a Fierz cases}
\fl J_{a}{}^{\mu}K_{0}{}^{\nu}+J_{0}{}^{\mu}K_{a}{}^{\nu}=(1/2)[\rmi J_{0}\Sdual_{a}{}^{\mu\nu}+\rmi J_{a}\Sdual_{0}{}^{\mu\nu}-\rmi K_{0}S_{a}{}^{\mu\nu}-\rmi K_{a}S_{0}{}^{\mu\nu}+J_{0}{}^{\mu}K_{a}{}^{\nu} \nonumber \\
+J_{0}{}^{\nu}K_{a}{}^{\mu}+J_{a}{}^{\mu}K_{0}{}^{\nu}+J_{a}{}^{\nu}K_{0}{}^{\mu}-J_{0\sigma}K_{a}{}^{\sigma}\eta^{\mu\nu}-J_{a\sigma}K_{0}{}^{\sigma}\eta^{\mu\nu}],
\end{eqnarray}
which eliminates the Pauli Levi-Civita contracted terms that are in the separate $a-0$ and $0-a$ cases, as they are identical and of opposite sign in each. Subtracting (\ref{Sum of a-0 and 0-a Fierz cases}) from itself, but with the $\mu, \nu$ terms flipped, will cancel all of the terms symmetric in $\mu, \nu$:
\begin{eqnarray}\label{Lorentz antisymmetric subtraction term, free Pauli index}
\fl(J_{a}{}^{\mu}K_{0}{}^{\nu}+J_{0}{}^{\mu}K_{a}{}^{\nu})-(J_{a}{}^{\nu}K_{0}{}^{\mu}+J_{0}{}^{\nu}K_{a}{}^{\mu}) \nonumber \\
=\rmi J_{0}\Sdual_{a}{}^{\mu\nu}+\rmi J_{a}\Sdual_{0}{}^{\mu\nu}-\rmi K_{0}S_{a}{}^{\mu\nu}-\rmi K_{a}S_{0}{}^{\mu\nu}.
\end{eqnarray}
Another way to get rid of the terms symmetric in $\mu, \nu$ is to contract (\ref{Sum of a-0 and 0-a Fierz cases}) with $\epsilon^{\mu\nu\rho\kappa}$:
\begin{eqnarray}
\fl\epsilon^{\mu\nu\rho\kappa}(J_{a\rho}K_{0\kappa}+J_{0\rho}K_{a\kappa})=(1/2)\epsilon^{\mu\nu\rho\kappa}[\rmi J_{0}\Sdual_{a\rho\kappa}+\rmi J_{a}\Sdual_{0\rho\kappa}-\rmi K_{0}S_{a\rho\kappa}-\rmi K_{a}S_{0\rho\kappa}] \nonumber \\
=-(1/4)(\epsilon^{\mu\nu\rho\kappa}\epsilon_{\rho\kappa\delta\tau})J_{0}S_{a}{}^{\delta\tau}-(1/4)(\epsilon^{\mu\nu\rho\kappa}\epsilon_{\rho\kappa\delta\tau})J_{a}S_{0}{}^{\delta\tau}-K_{0}\Sdual_{a}{}^{\mu\nu} \nonumber \\
\qquad-K_{a}\Sdual_{0}{}^{\mu\nu} \nonumber \\
=-(1/2)(\delta^{\mu}{}_{\tau}\delta^{\nu}{}_{\delta}-\delta^{\mu}{}_{\delta}\delta^{\nu}{}_{\tau})J_{0}S_{a}{}^{\delta\tau}-(1/2)(\delta^{\mu}{}_{\tau}\delta^{\nu}{}_{\delta}-\delta^{\mu}{}_{\delta}\delta^{\nu}{}_{\tau})J_{a}S_{0}{}^{\delta\tau} \nonumber \\
\qquad-K_{0}\Sdual_{a}{}^{\mu\nu}-K_{a}\Sdual_{0}{}^{\mu\nu},
\end{eqnarray}
where we have used the Levi-Civita tensor double contraction (\ref{Levi-Civita double contraction identity}). Simplifying gives
\begin{eqnarray}\label{Lorentz antisymmetric Levi-Civita contracted term, free Pauli index}
\fl\epsilon^{\mu\nu\rho\kappa}(J_{a\rho}K_{0\kappa}+J_{0\rho}K_{a\kappa})=J_{0}S_{a}{}^{\mu\nu}+J_{a}S_{0}{}^{\mu\nu}-K_{0}\Sdual_{a}{}^{\mu\nu}-K_{a}\Sdual_{0}{}^{\mu\nu}.
\end{eqnarray}
Now take the combination of (\ref{Lorentz antisymmetric subtraction term, free Pauli index}) and (\ref{Lorentz antisymmetric Levi-Civita contracted term, free Pauli index}):
\begin{eqnarray}\label{Sa Fierz identity before rearrangement and substitution}
\fl J_{0}\epsilon^{\mu\nu\rho\kappa}(J_{a\rho}K_{0\kappa}+J_{0\rho}K_{a\kappa})-\rmi K_{0}[(J_{a}{}^{\mu}K_{0}{}^{\nu}+J_{0}{}^{\mu}K_{a}{}^{\nu})-(J_{a}{}^{\nu}K_{0}{}^{\mu}+J_{0}{}^{\nu}K_{a}{}^{\mu})] \nonumber \\
=J_{0}^{2}S_{a}{}^{\mu\nu}+J_{0}J_{a}S_{0}{}^{\mu\nu}-J_{0}K_{0}\Sdual_{a}{}^{\mu\nu}-J_{0}K_{a}\Sdual_{0}{}^{\mu\nu}+K_{0}J_{0}\Sdual_{a}{}^{\mu\nu}+K_{0}J_{a}\Sdual_{0}{}^{\mu\nu} \nonumber \\
\qquad-K_{0}^{2}S_{a}{}^{\mu\nu}-K_{0}K_{a}S_{0}{}^{\mu\nu} \nonumber \\
=(J_{0}^{2}-K_{0}^{2})S_{a}{}^{\mu\nu}+(J_{0}J_{a}-K_{0}K_{a})S_{0}{}^{\mu\nu}+(K_{0}J_{a}-J_{0}K_{a})\Sdual_{0}{}^{\mu\nu}.
\end{eqnarray}
Let us rewrite the second and third terms on the right-hand side, by substituting the $S_{0}{}^{\mu\nu}$ and $\Sdual_{0}{}^{\mu\nu}$ identities, (\ref{S0 Fierz identity appendix}) and (\ref{Sdual0 Fierz identity appendix}):
\begin{eqnarray}
\fl(J_{0}J_{a}-K_{0}K_{a})S_{0}{}^{\mu\nu}=(1/2)(J_{0}^{2}-K_{0}^{2})^{-1}[(J_{0}^{2}J_{a}-J_{0}K_{0}K_{a})\epsilon^{\mu\nu\rho\kappa}J_{i\rho}K^{i}{}_{\kappa} \nonumber \\
+\rmi(K_{0}^{2}K_{a}-J_{0}K_{0}J_{a})(J_{i}{}^{\mu}K^{i\nu}-J_{i}{}^{\nu}K^{i\mu})].
\end{eqnarray}
Likewise, for the dual we have:
\begin{eqnarray}
\fl(K_{0}J_{a}-J_{0}K_{a})\Sdual_{0}{}^{\mu\nu}=(1/2)(J_{0}^{2}-K_{0}^{2})^{-1}[(K_{0}^{2}J_{a}-J_{0}K_{0}K_{a})\epsilon^{\mu\nu\rho\kappa}J_{i\rho}K^{i}{}_{\kappa} \nonumber \\
+\rmi(J_{0}^{2}K_{a}-J_{0}K_{0}J_{a})(J_{i}{}^{\mu}K^{i\nu}-J_{i}{}^{\nu}K^{i\mu})].
\end{eqnarray}
Summing these two equations together gives
\begin{eqnarray}
\fl(J_{0}J_{a}-K_{0}K_{a})S_{0}{}^{\mu\nu}+(K_{0}J_{a}-J_{0}K_{a})\Sdual_{0}{}^{\mu\nu} \nonumber \\
=\frac{J_{0}^{2}+K_{0}^{2}}{2(J_{0}^{2}-K_{0}^{2})}[J_{a}\epsilon^{\mu\nu\rho\kappa}J_{i\rho}K^{i}{}_{\kappa}+\rmi K_{a}(J_{i}{}^{\mu}K^{i\nu}-J_{i}{}^{\nu}K^{i\mu})] \nonumber \\
\qquad-\frac{J_{0}K_{0}}{J_{0}^{2}-K_{0}^{2}}[K_{a}\epsilon^{\mu\nu\rho\kappa}J_{i\rho}K^{i}{}_{\kappa}+\rmi J_{a}(J_{i}{}^{\mu}K^{i\nu}-J_{i}{}^{\nu}K^{i\mu})].
\end{eqnarray}
Substituting this into (\ref{Sa Fierz identity before rearrangement and substitution}) and rearranging, we finally get
\begin{eqnarray}\label{Sa Fierz identity appendix}
\fl S_{a}{}^{\mu\nu}=(J_{0}^{2}-K_{0}^{2})^{-1}\{J_{0}\epsilon^{\mu\nu\rho\kappa}(J_{a\rho}K_{0\kappa}+J_{0\rho}K_{a\kappa}) \nonumber \\
-\rmi K_{0}[(J_{a}{}^{\mu}K_{0}{}^{\nu}+J_{0}{}^{\mu}K_{a}{}^{\nu})-(J_{a}{}^{\nu}K_{0}{}^{\mu}+J_{0}{}^{\nu}K_{a}{}^{\mu})]\} \nonumber \\
-\frac{J_{0}^{2}+K_{0}^{2}}{2(J_{0}^{2}-K_{0}^{2})^{2}}[J_{a}\epsilon^{\mu\nu\rho\kappa}J_{i\rho}K^{i}{}_{\kappa}+\rmi K_{a}(J_{i}{}^{\mu}K^{i\nu}-J_{i}{}^{\nu}K^{i\mu})] \nonumber \\
+\frac{J_{0}K_{0}}{(J_{0}^{2}-K_{0}^{2})^{2}}[K_{a}\epsilon^{\mu\nu\rho\kappa}J_{i\rho}K^{i}{}_{\kappa}+\rmi J_{a}(J_{i}{}^{\mu}K^{i\nu}-J_{i}{}^{\nu}K^{i\mu})].
\end{eqnarray}
Now let us derive the dual, $\Sdual_{a}{}^{\mu\nu}$ by considering the alternate combination:
\begin{eqnarray}\label{Sduala Fierz identity before rearrangement and substitution}
\fl K_{0}\epsilon^{\mu\nu\rho\kappa}(J_{a}{}_{\rho}K_{0\kappa}+J_{0\rho}K_{a}{}_{\kappa})-\rmi J_{0}[(J_{a}{}^{\mu}K_{0}{}^{\nu}+J_{0}{}^{\mu}K_{a}{}^{\nu})-(J_{a}{}^{\nu}K_{0}{}^{\mu}+J_{0}{}^{\nu}K_{a}{}^{\mu})] \nonumber \\
=J_{0}K_{0}S_{a}{}^{\mu\nu}+K_{0}J_{a}S_{0}{}^{\mu\nu}-K_{0}^{2}\Sdual_{a}{}^{\mu\nu}-K_{0}K_{a}\Sdual_{0}{}^{\mu\nu}+J_{0}^{2}\Sdual_{a}{}^{\mu\nu}+J_{0}J_{a}\Sdual_{0}{}^{\mu\nu} \nonumber \\
\qquad-J_{0}K_{0}S_{a}{}^{\mu\nu}-J_{0}K_{a}S_{0}{}^{\mu\nu} \nonumber \\
=(J_{0}^{2}-K_{0}^{2})\Sdual_{a}{}^{\mu\nu}+(K_{0}J_{a}-J_{0}K_{a})S_{0}{}^{\mu\nu}+(J_{0}J_{a}-K_{0}K_{a})\Sdual_{0}{}^{\mu\nu}.
\end{eqnarray}
Again, we rewrite the second and third terms, by substituting identities (\ref{S0 Fierz identity appendix}) and (\ref{Sdual0 Fierz identity appendix}):
\begin{eqnarray}
\fl(K_{0}J_{a}-J_{0}K_{a})S_{0}{}^{\mu\nu}=(1/2)(J_{0}^{2}-K_{0}^{2})^{-1}[(J_{0}K_{0}J_{a}-J_{0}^{2}K_{a})\epsilon^{\mu\nu\rho\kappa}J_{i\rho}K^{i}{}_{\kappa} \nonumber \\
+\rmi(J_{0}K_{0}K_{a}-K_{0}^{2}J_{a})(J_{i}{}^{\mu}K^{i\nu}-J_{i}{}^{\nu}K^{i\mu})].
\end{eqnarray}
The dual term is
\begin{eqnarray}
\fl(J_{0}J_{a}-K_{0}K_{a})\Sdual_{0}{}^{\mu\nu}=(1/2)(J_{0}^{2}-K_{0}^{2})^{-1}[(J_{0}K_{0}J_{a}-K_{0}^{2}K_{a})\epsilon^{\mu\nu\rho\kappa}J_{i\rho}K^{i}{}_{\kappa} \nonumber \\
+\rmi(J_{0}K_{0}K_{a}-J_{0}^{2}J_{a})(J_{i}{}^{\mu}K^{i\nu}-J_{i}{}^{\nu}K^{i\mu})].
\end{eqnarray}
Summing the two:
\begin{eqnarray}
\fl(K_{0}J_{a}-J_{0}K_{a})S_{0}{}^{\mu\nu}+(J_{0}J_{a}-K_{0}K_{a})\Sdual_{0}{}^{\mu\nu} \nonumber \\
=\frac{J_{0}K_{0}}{J_{0}^{2}-K_{0}^{2}}[J_{a}\epsilon^{\mu\nu\rho\kappa}J_{i\rho}K^{i}{}_{\kappa}+\rmi K_{a}(J_{i}{}^{\mu}K^{i\nu}-J_{i}{}^{\nu}K^{i\mu})] \nonumber \\
\qquad-\frac{(J_{0}^{2}+K_{0}^{2})}{2(J_{0}^{2}-K_{0}^{2})}[K_{a}\epsilon^{\mu\nu\rho\kappa}J_{i\rho}K^{i}{}_{\kappa}+\rmi J_{a}(J_{i}{}^{\mu}K^{i\nu}-J_{i}{}^{\nu}K^{i\mu})].
\end{eqnarray}
Finally, after substituting this into (\ref{Sduala Fierz identity before rearrangement and substitution}) and rearranging, we get:
\begin{eqnarray}
\fl\Sdual_{a}{}^{\mu\nu}=(J_{0}^{2}-K_{0}^{2})^{-1}\{K_{0}\epsilon^{\mu\nu\rho\kappa}(J_{a\rho}K_{0\kappa}+J_{0\rho}K_{a\kappa}) \nonumber \\
-\rmi J_{0}[(J_{a}{}^{\mu}K_{0}{}^{\nu}+J_{0}{}^{\mu}K_{a}{}^{\nu})-(J_{a}{}^{\nu}K_{0}{}^{\mu}+J_{0}{}^{\nu}K_{a}{}^{\mu})]\} \nonumber \\
+\frac{J_{0}^{2}+K_{0}^{2}}{2(J_{0}^{2}-K_{0}^{2})^{2}}[K_{a}\epsilon^{\mu\nu\rho\kappa}J_{i\rho}K^{i}{}_{\kappa}+\rmi J_{a}(J_{i}{}^{\mu}K^{i\nu}-J_{i}{}^{\nu}K^{i\mu})] \nonumber \\
-\frac{J_{0}K_{0}}{(J_{0}^{2}-K_{0}^{2})^{2}}[J_{a}\epsilon^{\mu\nu\rho\kappa}J_{i\rho}K^{i}{}_{\kappa}+\rmi K_{a}(J_{i}{}^{\mu}K^{i\nu}-J_{i}{}^{\nu}K^{i\mu})].
\end{eqnarray}
Comparing this identity with the identity for $S_{a}{}^{\mu\nu}$, (\ref{Sa Fierz identity appendix}), we can make the observations that the first terms are the same in each, but with the $J_{0}$ and $K_{0}$ terms interchanged (similarly to the $S_{0}{}^{\mu\nu}$ and $\Sdual_{0}{}^{\mu\nu}$ comparison). Likewise, the second and third terms of both identities are the same, but with the $J_{a}$ and $K_{a}$ terms interchanged, \textit{along with the signs}, which are flipped with respect to each other.


\Bibliography{9}

\bibitem{Booth-2000} Booth H 2002 Nonlinear electron solutions and their characterisations at infinity {\it ANZIAM J.} {\bf 44} 51--59

\bibitem{Flato-Simon-Taflin-1997} Flato M, Simon J C H, Taflin E 1997 Asymptotic completeness, global existence and the infrared problem for the Maxwell-Dirac equations {\it Memoirs of the Amer. Math. Soc.} {\bf 127} 606:1--312

\bibitem{Radford-1996} Radford C J 1996 Localized solutions of the Dirac-Maxwell equations {\it \JMP} {\bf 37} 4418--33

\bibitem{Booth-Radford-1997} Booth H S and Radford C J 1997 The Dirac-Maxwell equations with cylindrical symmetry {\it \JMP} {\bf 38} 1257--68

\bibitem{Radford-Booth-1999} Radford C and Booth H 1999 Magnetic monopoles, electric neutrality and the static Maxwell-Dirac equations {\it \JPA} {\bf 32} 5807--22

\bibitem{Legg-2007} Legg G P 2007 On Group invariant solutions to the Maxwell Dirac equations {\it M.Sc. Thesis: University of Tasmania}

\bibitem{Eliezer-1958} Eliezer C J 1958 A consistency condition for electron wave functions {\it Proc. Cam. Phil. Soc.} {\bf 54} 247

\bibitem{Booth-Legg-Jarvis-2001} Booth H S, Legg G and Jarvis P D 2001 Algebraic solution for the vector potential in the Dirac equation {\it \JPA} {\bf 34} 5667--77

\bibitem{Takabayasi-1957} Takabayasi T 1957 Relativistic hydrodynamics of the Dirac matter {\it Prog. Theor. Phys. Suppl.} {\bf 4} 2--80

\bibitem{Bistrovic-2003} Bistrovic B, Jackiw R, Li H, Nair V P and Pi S-Y 2003 Non-abelian fluid dynamics in Lagrangian formulation {\it Phys. Rev. D} {\bf 67} 025013

\bibitem{Crawford-1985} Crawford J P 1985 On the algebra of Dirac bispinor densities: Factorization and inversion theorems {\it \JMP} {\bf 26} 1439--41

\bibitem{Itzykson-Zuber-1980} Itzykson C and Zuber J-P 1980 {\it Quantum field theory} (New York: McGraw-Hill International Book Co.) p 85

\bibitem{Hogben-2007} Hogben L (ed) 2007 {\it Handbook of Linear Algebra} ({\it Discrete Mathematics and its Applications}) ed R Brualdi, A Greenbaum, R Mathias (Boca Raton: Chapman \& Hall/CRC, Taylor \& Francis Group) p~{\bf 14}--16

\bibitem{Delbourgo-Prasad-1974} Delbourgo R and Prasad V B 1974 Spinor interactions in the two-dimensional limit {\it Nuovo Cimento A Serie} {\bf 21} 32--44

\bibitem{Kaempffer-1981} Kaempffer F A 1981 Spinor electrodynamics as a dynamics of currents {\it Phys. Rev. D} {\bf 23} 918--21

\bibliographystyle{unsrt}


\endbib
\end{document}